\documentclass[sigconf]{aamas}

\usepackage{balance} 
\usepackage{tabularx}
\usepackage{stfloats}

\usepackage{booktabs}
\usepackage{multirow}
\usepackage{adjustbox}
\usepackage{threeparttable}

\setcopyright{ifaamas}
\acmConference[AAMAS '26]{Proc.\@ of the 25th International Conference
on Autonomous Agents and Multiagent Systems (AAMAS 2026)}{May 25 -- 29, 2026}
{Paphos, Cyprus}{C.~Amato, L.~Dennis, V.~Mascardi, J.~Thangarajah (eds.)}
\copyrightyear{2026}
\acmYear{2026}
\acmDOI{}
\acmPrice{}
\acmISBN{}

\acmSubmissionID{18}

\title[Strategic Persuasion with Trait-Conditioned Multi-Agent Systems for Iterative Legal Argumentation]{Strategic Persuasion with Trait-Conditioned\\Multi-Agent Systems for Iterative Legal Argumentation}

\author{Philipp D. Siedler}
\authornote{Secondary email: p.d.siedler@gmail.com}
\affiliation{
  \institution{Aleph Alpha Research}
  \city{Heidelberg}
  \country{Germany}}
\email{philipp.siedler@aleph-alpha-research.com}

\begin{abstract}
Strategic interaction in adversarial domains such as law, diplomacy, and negotiation is mediated by language, yet most game-theoretic models abstract away the mechanisms of persuasion that operate through discourse. We present the Strategic Courtroom Framework, a multi-agent simulation environment in which prosecution and defense teams composed of trait-conditioned Large Language Model (LLM) agents engage in iterative, round-based legal argumentation. Agents are instantiated using nine interpretable traits organized into four archetypes, enabling systematic control over rhetorical style and strategic orientation.

We evaluate the framework across 10 synthetic legal cases and 84 three-trait team configurations, totaling over 7{,}000 simulated trials using DeepSeek-R1 and Gemini~2.5~Pro. Our results show that heterogeneous teams with complementary traits consistently outperform homogeneous configurations, that moderate interaction depth yields more stable verdicts, and that certain traits (notably quantitative and charismatic) contribute disproportionately to persuasive success. We further introduce a reinforcement-learning-based Trait Orchestrator that dynamically generates defense traits conditioned on the case and opposing team, discovering strategies that outperform static, human-designed trait combinations.

Together, these findings demonstrate how language can be treated as a first-class strategic action space and provide a foundation for building autonomous agents capable of adaptive persuasion in multi-agent environments.
\end{abstract}

      
\newcommand{\BibTeX}{\rm B\kern-.05em{\sc i\kern-.025em b}\kern-.08em\TeX}

\begin{document}

\pagestyle{fancy}
\fancyhead{}

\maketitle 

\section{Introduction} \label{intro}

Strategic interaction in the real world is often mediated by language. In adversarial discourse settings -- such as legal proceedings, diplomatic bargaining, and organizational conflict -- participants do not merely choose actions with explicit payoffs; they craft narratives, contest facts, and deploy rhetorical strategies to shape an adjudicator's belief state. Classical game-theoretic models capture incentives and equilibrium structure under well-defined action spaces, but they typically abstract away the linguistic and social mechanisms by which persuasion unfolds.

Large Language Models (LLMs) offer a complementary modeling substrate. Because LLMs can generate coherent arguments conditioned on context and style, they enable agent-based simulations in which strategy is expressed through language rather than discrete action primitives. This opens the door to studying questions that are difficult to formalize: how personality-driven rhetoric interacts with evidence, how diverse teams coordinate persuasive labor, and how iterative exchange changes outcomes over time.

In this paper, we introduce the \textbf{Strategic Courtroom Framework}, a multi-agent simulation environment for adversarial legal argumentation. The framework instantiates prosecution and defense as heterogeneous teams of LLM-based agents that engage in a structured, multi-round debate over a case record (summary, evidence, and legal issues). A judge evaluates the competing arguments and returns a discrete verdict along with a confidence score. This setup yields a controllable testbed for analyzing how \emph{team composition}, \emph{interaction depth}, and \emph{agent traits} shape persuasive outcomes.

Our contributions are:

\begin{enumerate}
    \item \textbf{Aristotelian Trait Taxonomy}: We define a taxonomy of nine traits organized into four interpretable archetypes (Rhetoricians, Technicians, Gladiators, and Diplomats), providing a principled basis for trait-driven agent design.
    
    \item \textbf{Iterative Argumentation Protocol}: We implement a multi-round adversarial protocol in which agents condition their arguments on the opponent's prior moves, enabling adaptive and path-dependent discourse dynamics.
    
    \item \textbf{Comprehensive Empirical Analysis}: We evaluate across 10 synthetic legal cases and 84 three-trait team configurations, totaling 7{,}000+ simulated trials using DeepSeek-R1 and Gemini 2.5 Pro, and analyze how traits, rounds, and team structure affect outcomes.
    
    \item \textbf{RL-based Trait Orchestration}: We introduce a reinforcement learning approach that fine-tunes a lightweight LLM-based orchestrator to generate defense traits conditioned on the case and opposing team, discovering strategies beyond a fixed trait library.
\end{enumerate}

Overall, our experiments show that heterogeneous teams with complementary traits can outperform homogeneous configurations, and that learned orchestration can further improve performance by adapting trait design to the adversary and case context.

\begin{table*}[b]
	\caption{Strategic Profiles and Behavioral Characteristics Based on Aristotelian Philosophy}
	\label{tab:profiles}
	\centering
	\begin{tabular*}{\textwidth}{@{\extracolsep{\fill}}llll}\toprule
		\textit{Archetype} & \textit{Trait} & \textit{Philosophy} & \textit{Behavior} \\ \midrule
		Rhetorician & Charismatic & Pathos & Appeals to emotions and rapport to sway judgment beyond mere facts \\
		Rhetorician & Folksy & Social Virtue & Appears as a peer to foster trust with the jury \\
		Rhetorician & Moralistic & Ethics & Frames the case through the lens of ``The Good'' \\
		\midrule
		Technician & Pedantic & Excess of Exactness & Extreme focus on the letter of the law \\
		Technician & Quantitative & Logos & Relies on logical demonstration and hard data \\
		\midrule
		Gladiator & Tenacious & Courage & Persists in difficult arguments despite pressure \\
		Gladiator & Provocative & Irascibility & Deliberately stirs conflict for tactical advantage \\
		\midrule
		Diplomat & Transparent & Truthfulness & Presents the case exactly as it is \\
		Diplomat & Methodical & Phronesis & Uses practical wisdom to guide through complexity \\ \bottomrule
	\end{tabular*}
\end{table*}

\subsection{Research Questions}

We investigate the following research questions:

\begin{itemize}
\item \textbf{RQ1}: Does trait diversity in multi-agent teams improve persuasive performance compared to homogeneous teams?
\item \textbf{RQ2}: Which individual traits and archetypes contribute most to persuasive success?
\item \textbf{RQ3}: How does iterative depth affect stability and quality of judicial outcomes?
\item \textbf{RQ4}: Can a learned trait orchestrator outperform static, human-designed trait combinations?
\end{itemize}

We hypothesize that heterogeneous teams outperform homogeneous teams (H1), that emotionally grounded traits, particularly \textit{charismatic}, exhibit consistent positive effects (H2), and that learned orchestration discovers strategies beyond predefined trait spaces (H3).

\begin{figure*}
    \centering
    \includegraphics[width=1\linewidth]{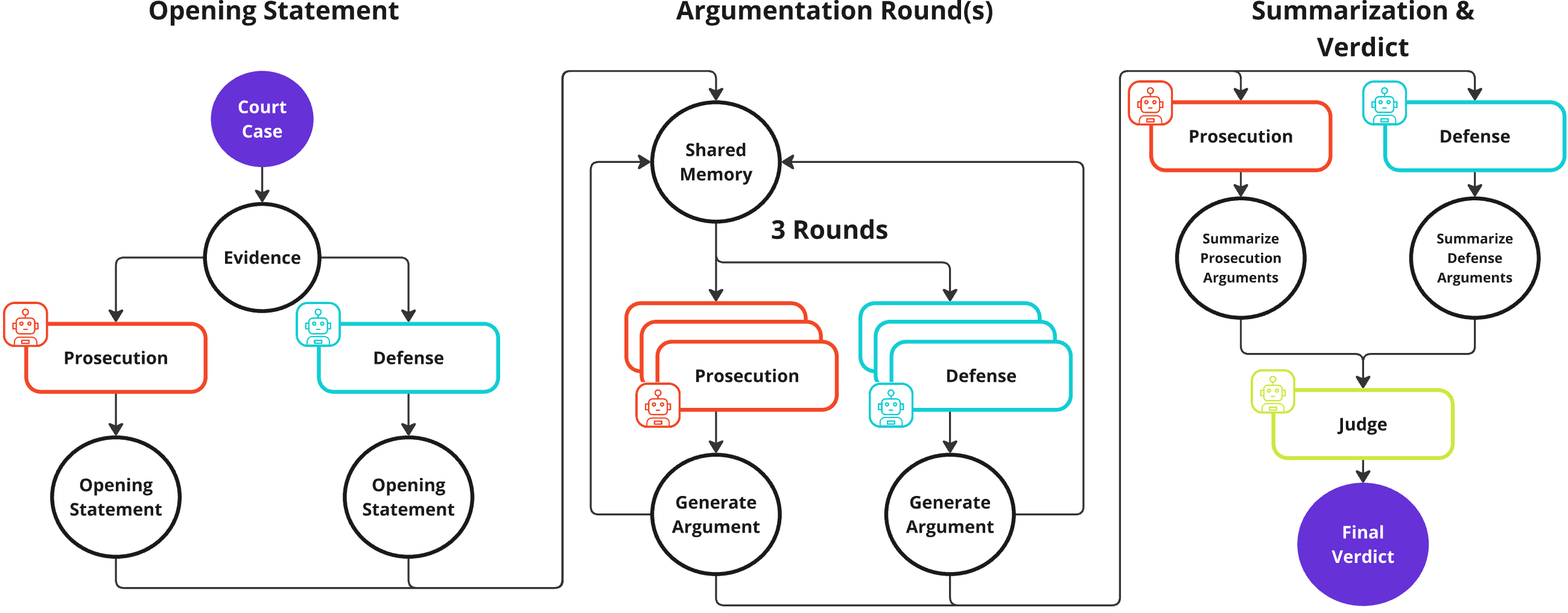}
    \caption{Overview of the Strategic Courtroom Framework. Prosecution and defense teams generate opening statements, engage in multi-round argumentation with shared memory, summarize positions, and a judge produces a final verdict.}
    \label{fig:thumbnail}
\end{figure*}

\begin{table}[H]
\centering
\caption{Mapping between research questions, evaluation metrics, and supporting evidence.}
\begin{adjustbox}{width=\columnwidth}
\begin{tabular}{lll}
\hline
Research Question & Primary Metrics & Evidence in Paper \\
\hline
RQ1: Trait diversity vs. homogeneity & Avg. Elo, Win rate & Table \ref{tab:aggregate} \\
RQ2: Individual trait importance & Trait frequency, Trait Elo & Tables \ref{tab:overall}--\ref{tab:bestconfigs} \\
RQ3: Effect of iterative depth & Verdict reversal rate, Avg. Elo by rounds & Section \ref{sec:round-depth-stability} \\
RQ4: Learned vs. static traits & Win rate vs. static best, Avg. Elo & Section \ref{sec:training-trait-orchestrator} \\
\hline
\end{tabular}
\end{adjustbox}
\end{table}

We organize Section 5 around these questions.

\section{Background \& Related Work}

\subsection{Multi-Agent Systems and LLMs}

Multi-agent systems (MAS) have long been studied in artificial intelligence for modeling complex interactions among autonomous entities~\cite{wooldridge_introduction_2009}. Recent work has explored using LLMs as the cognitive backbone of agents, enabling natural language communication and reasoning~\cite{park_generative_2023}. Systems like AutoGen~\cite{wu_autogen_2023} and MetaGPT~\cite{hong_metagpt_2023} demonstrate that LLM-based agents can collaborate on complex tasks, though primarily in cooperative rather than adversarial settings.

\subsection{Computational Argumentation}

The field of computational argumentation has developed formal frameworks for modeling debate and persuasion~\cite{simari_argumentation_2009}. Argument mining techniques extract argumentative structures from text~\cite{lawrence_argument_2019}, while argumentation schemes provide templates for common reasoning patterns~\cite{visser_annotating_2021}. Our work complements these formal approaches by using LLMs to generate contextually appropriate arguments within a strategic game framework.

\subsection{Persuasion, Signaling, and Rhetorical Games}

Our work is also related to a large body of game-theoretic research on persuasion and information transmission. Bayesian Persuasion studies how a sender strategically designs information structures to influence a receiver’s beliefs and actions under asymmetric information \cite{kamenica_bayesian_2011}. Subsequent work has examined optimal signaling, commitment, and robustness properties of such mechanisms, typically assuming abstract signal spaces and rational receivers.

Signaling games and cheap-talk models analyze how communication can convey private information even when messages are costless, capturing conditions under which informative equilibria emerge \cite{crawford_strategic_1982}. Extensions of these models consider richer message spaces and boundedly rational agents, but still largely treat language as symbolic tokens rather than natural language utterances.

Rhetorical and debate-style games further explore strategic argumentation, where agents choose argumentative moves to affect beliefs or verdicts \cite{amgoud_using_2009,simari_argumentation_2009}. These frameworks provide formal semantics and acceptability criteria for arguments, but typically rely on predefined argument structures.

In contrast to these lines of work, our framework treats free-form natural language generated by LLM agents as the primary strategic action space. Rather than solving for equilibrium policies in a handcrafted game, we empirically study how persona-conditioned agents interact, adapt, and persuade through discourse. This positions our approach as complementary to classical persuasion and signaling models, offering a simulation-based testbed for exploring strategic communication in high-dimensional linguistic spaces.

\subsection{AI in Legal Reasoning}

Legal AI has advanced significantly with LLMs demonstrating competence on bar examinations~\cite{katz_gpt-4_2023} and legal document analysis~\cite{chalkidis_legal-bert_2020}. Prior work on legal prediction focuses on case outcome forecasting~\cite{noauthor_using_nodate}, while we focus on the strategic dynamics of adversarial argumentation. Recent systems have explored judge simulation~\cite{hamilton_blind_2023}, but few examine the full dynamics of multi-party legal proceedings with heterogeneous agent profiles.

\subsection{Trait-Based Language Generation}

Trait-conditioned text generation has shown that LLMs can maintain consistent behavioral patterns when prompted appropriately~\cite{li_persona-based_2016,zhang_personalizing_2018}. We extend this work by grounding traits in Aristotelian virtue ethics, providing a principled taxonomy rather than ad-hoc personality descriptions, and studying how trait combinations interact in adversarial settings.

\section{The Strategic Courtroom Framework}

Our architecture decomposes a legal case into a multi-stage strategic game where prosecution and defense teams compete to influence a judge through iterative argumentation. We also provide a game-theoretic interpretation of this environment to clarify its relationship to classical multi-agent models.

\subsection{Court Case Data}

We construct a corpus of 10 synthetic legal cases spanning diverse legal domains: assault (self-defense claims), breach of contract, petty theft, defamation, eviction disputes, DUI charges, medical malpractice, noise violations, employment discrimination, and personal injury. Each case is represented as a structured tuple:

\begin{equation}
    \mathcal{C} = \langle \text{name}, \text{summary}, \mathcal{E}, \mathcal{I} \rangle
\end{equation}

\noindent where $\mathcal{E} = \{e_1, ..., e_k\}$ is a set of evidence items (e.g., ``Security camera footage,'' ``Medical records'') and $\mathcal{I} = \{i_1, ..., i_m\}$ is a set of legal issues (e.g., ``self-defense,'' ``intent to steal''). This structured representation allows agents to ground their arguments in specific factual claims while addressing relevant legal standards.

All cases in our corpus are synthetically generated using an LLM and subsequently manually sanity-checked by the authors to ensure internal consistency, plausible fact patterns, and coverage of the stated legal issues. We adopt synthetic cases intentionally: our goal is not to model real-world legal doctrine in full fidelity, but to isolate and study persuasion mechanisms, interaction dynamics, and trait effects under controlled conditions. Synthetic cases allow systematic variation of evidence structure and issue composition while avoiding confounds introduced by jurisdiction-specific rules, incomplete records, or latent real-world correlations. As such, the cases should be viewed as abstract task environments for mechanism discovery rather than substitutes for real court proceedings.

\subsection{System Architecture}

The system consists of three core components:

\textbf{The Case Environment}: A structured input containing evidence, summary, and legal issues that defines the factual and legal landscape of the dispute.

\textbf{Strategic Teams}: Heterogeneous groups of agents (Prosecution/Defense) that rotate turns to build a coherent narrative. Each team consists of $n$ agents (typically $n=3$), each instantiated with a distinct character trait that shapes their rhetorical approach. In team mode, agents take turns contributing arguments in round-robin fashion, enabling ``emergent strategy'' where one agent's approach sets up another's.

\textbf{The Judge}: A ``Judge'' agent tasked with evaluating the competitive arguments and providing a machine-readable verdict. The judge receives summarized arguments from both sides and produces:
\begin{equation}
    V = \langle v \in \{\text{guilty}, \text{not guilty}\}, c \in [0,1] \rangle
\end{equation}
where $v$ is the verdict and $c$ is a confidence score reflecting the judge's certainty.

\subsection{Iterative Argumentation Protocol}

Unlike single-shot prompting approaches, our agents engage in $N$ rounds of structured debate. The protocol proceeds as follows:

\textbf{Opening Statements}: Both prosecution and defense generate initial position statements summarizing their case theory.\\
\textbf{Argument Rounds}: For each round $r \in \{1, ..., N\}$ and each legal issue $i \in \mathcal{I}$: Prosecution generates argument $a_p^{r,i}$ conditioned on defense's previous argument, then the defense generates rebuttal $a_d^{r,i}$ conditioned on prosecution's argument.\\
\textbf{Deliberation}: Each team summarizes their cumulative arguments, and the judge renders a verdict based on these summaries.

Each agent's ``Strategic World Model'' is updated by incorporating the preceding argument from the opposing side into their prompt context, creating a dynamic tree of claims and counter-claims that enables adaptive argumentation.

\subsection{Game-Theoretic Interpretation}

The Strategic Courtroom Framework can be viewed as a sequential, partially observable stochastic game in which language constitutes the primary action space. At any timestep, the environment state consists of the case specification together with the history of arguments exchanged so far:
\[
s_t = \langle C, h_t \rangle,
\]
where $C$ is the structured case representation (summary, evidence, legal issues) and $h_t$ is the discourse history up to time $t$.

Each agent’s action is a natural-language utterance conditioned on its role, traits, and the current state:
\[
a_t \in \mathcal{U},
\]
where $\mathcal{U}$ denotes the space of possible textual arguments.

After a fixed number of rounds, the judge maps the final state to an outcome and confidence score,
\[
V = \langle v, c \rangle,
\]
which induces terminal payoffs for the prosecution and defense (win/loss). In this view, persona-conditioned prompting defines a restricted policy class over utterances, and learning or selecting traits corresponds to searching over high-level policy priors rather than directly optimizing token-level policies.

This interpretation situates our framework within the tradition of multi-agent sequential decision-making while accommodating high-dimensional natural language actions.

\subsection{Agent Implementation}

Each \texttt{CourtAgent} is instantiated with a role, case data, trait list, and LLM backend. The agent's system prompt encodes its personality:

\begin{quote}
\textit{``You are a [traits] [role] Agent in a court case. Your role is to contribute to the trial by providing arguments based on the context of the case. Adopt a tone that reflects your personality as a [traits] [role]. Be super concise.''}
\end{quote}

The \texttt{Team} class manages agent rotation, ensuring each team member contributes in sequence. This design enables studying both individual agent effectiveness and emergent team dynamics.

\subsection{Judge and Evaluation Protocol}

Each simulated trial includes a dedicated Judge agent responsible for evaluating the prosecution and defense arguments and producing a final verdict. For experimental consistency, we use a single LLM backend per experiment: the same model instance is used for all courtroom agents (prosecution, defense, and judge) within that experimental condition. Thus, in DeepSeek-R1 experiments, all agents -- including the judge -- are instantiated with DeepSeek-R1, and analogously for Gemini-2.5-Pro.

The Judge agent is not trait-conditioned in the same manner as argumentative agents. Instead, its behavior is governed by a fixed, hardcoded role description emphasizing neutrality and procedural fairness. Concretely, the judge is prompted with the traits \textit{fair} and \textit{ethical}, which instruct the model to weigh evidence impartially, avoid rhetorical style preferences, and base decisions on the coherence and support of the presented arguments rather than their emotional tone.

All agents in the framework, including the judge, share a common system prompt template. The only difference across agents is the role string and the trait list. Concretely, the judge is instantiated with role = \textit{judge} and traits = \textit{fair, ethical}, yielding the following system prompt:

\textit{"You are an fair, ethical Judge Agent in a court case. Your role is to contribute to the trial by providing arguments, responses, or decisions based on the context of the case. Adopt a tone that reflects your personality as an fair, ethical judge. Be super concise."}

The judge receives summarized arguments from both sides at the end of the debate and outputs a structured decision:
\[
V = \langle v \in \{\text{guilty}, \text{not guilty}\}, c \in [0,1] \rangle
\]
where $v$ is the verdict and $c$ is a confidence score reflecting the judge's internal certainty.

All judge generations use the same decoding parameters as the argumentative agents within a given experiment (temperature = 0.7, top-p = 0.9, maximum tokens = 512). This design isolates the effect of trait-conditioned argumentation while avoiding confounds introduced by heterogeneous evaluation models.

Because the judge shares the same underlying model as the advocates, performance differences should be interpreted as arising from interaction dynamics and trait composition rather than cross-model evaluation artifacts. We further assess judge stability via re-evaluation with independently seeded judge instances (Section~\ref{judge-reli}).

\section{Design of Agents and Strategic Profiles}

A key contribution of this work is the use of \textbf{Trait-Driven Strategic Engineering}. We hypothesize that a ``mixture of personalities'' outperforms a monolithic approach in persuasive tasks, as different traits provide coverage across distinct dimensions of persuasion.

\subsection{Aristotelian Trait Taxonomy}

For brevity, we provide concise trait summaries here and full definitions in Appendix~\ref{app:traits}.

We ground our trait design in Aristotelian philosophy, organizing 9 traits into four archetypes that correspond to classical modes of persuasion and virtue:

Importantly, our use of Aristotelian terminology is not intended as a faithful reconstruction of Aristotelian psychology. Rather, Aristotelian concepts serve as an organizing metaphor that provides interpretable anchors for different persuasive tendencies (emotional appeal, logical rigor, persistence, and practical judgment). The traits are operationalized entirely through prompt-level behavioral constraints, and their effectiveness is evaluated empirically.

\textbf{Rhetoricians} ($\rho$): Agents focused on Aristotle's \textit{Ethos} and \textit{Pathos}- establishing credibility and emotional connection.
\textbf{Technicians} ($\tau$): Agents emphasizing \textit{Logos} and \textit{Akribeia} (exactness) -- logical rigor and precision.
\textbf{Gladiators} ($\gamma$): Agents exhibiting \textit{Andreia} (courage/spirit) -- persistence and confrontation.
\textbf{Diplomats} ($\delta$): Agents practicing \textit{Phronesis} (practical wisdom) -- measured judgment and transparency. Description of traits, philosophy and behavior can be found in Table \ref{tab:profiles}.

\subsection{Combinatorial Space}

With 9 traits and teams of size 3, we have $\binom{9}{3} = 84$ unique trait combinations per team. When considering ordered permutations (where trait assignment order to agents matters), this expands to $P(9,3) = 504$ configurations. Our experiments explore this combinatorial space systematically to identify effective strategies.

\subsection{Implementation Details}

Using the \texttt{CourtAgent} and \texttt{Team} classes, we instantiate these profiles through trait-specific system prompts. Each agent receives instructions encoding its archetype's behavioral tendencies, enabling observation of ``emergent strategy'' where, for example, a \textit{provocative} prosecutor destabilizes the defense, allowing a \textit{quantitative} colleague to substantiate claims with evidence while the opposition is off-balance.

\subsection{Proposed Experiments}

To evaluate the Strategic Engineering capabilities of this system, we propose three primary experiments:

\subsubsection{Experiment 1: The "Diversity Dividend"}

\textbf{Setup}: Compare a team of 3 "Logical" agents against a team of 1 "Aggressive," 1 "Logical," and 1 "Empathetic" agent.
\textbf{Metric}: Measure the "Win Rate" (verdicts in favor) and the Judge’s confidence score.
\textbf{Goal}: Determine if heterogeneous teams cover more "strategic ground."

\subsubsection{Experiment 2: Sensitivity to Judicial Bias}

\textbf{Setup}: Run the same trial evidence through three different Judge profiles: Strict Legalist, Compassionate Reformer, and Unbiased Machine.
\textbf{Metric}: Delta in confidence scores and verdict shifts.
\textbf{Goal}: Map how "Strategic Engineering" must account for the perceptions of the decision-maker.

\subsubsection{Experiment 3: Argumentative Decay vs. Depth}

\textbf{Setup}: Scale the number of rounds from 1 to 10.
\textbf{Metric}: Argument quality (using a secondary LLM-as-a-judge) and repetition rate.
\textbf{Goal}: Identify the point where iterative strategic reasoning hits diminishing returns.

\subsection{Experiment 4: RL-based Trait Orchestration}

\textbf{Hypothesis}: A learned orchestrator can discover novel trait combinations that outperform exhaustive search over predefined traits.

\textbf{Setup}: We train a Qwen2.5-1.5B-Instruct model using REINFORCE policy gradient to generate defense team traits given case information and prosecution traits. The orchestrator is not limited to predefined traits and can invent novel traits.

\textbf{Reward Function}:
\begin{equation}
    R(v, c) = \begin{cases}
        +c & \text{if } v = \text{``not guilty''} \\
        -c & \text{if } v = \text{``guilty''} \\
        0 & \text{otherwise}
    \end{cases}
\end{equation}

\textbf{Training}: We use LoRA fine-tuning (rank 16, $\alpha=32$) with 4-bit quantization for efficiency. The model is trained for 500 episodes with a learning rate search over $\{10^{-5}, 5 \times 10^{-5}, 10^{-4}\}$.

\section{Results}

We treat each simulated trial as a match between prosecution and defense and compute Elo ratings accordingly. Ratings are initialized at 1500 and updated using standard Elo update rules with $K=32$.

Tables~\ref{tab:prosecution}--\ref{tab:aggregate} summarize performance across experimental dimensions. We organize our analysis around the research questions posed in Section~\ref{intro}.

\paragraph{Elo computation (trait-level) and pooling.}
We compute Elo ratings at the \emph{trait} level by treating each trial as a match between the prosecution’s trait set and the defense’s trait set. For a given trial, we define each side’s rating as the mean Elo of its constituent traits. Let $\bar{R}_P$ and $\bar{R}_D$ be the prosecution and defense mean ratings, respectively. The expected score for the defense is
\[
E_D = \frac{1}{1 + 10^{(\bar{R}_P - \bar{R}_D)/400}},
\]
with $E_P = 1 - E_D$. Observed scores are determined from the verdict: for \textit{not guilty}, $(S_D,S_P)=(1,0)$; for \textit{guilty}, $(S_D,S_P)=(0,1)$; and for \textit{undecided}, $(S_D,S_P)=(0.5,0.5)$ (draw). Each trait appearing in the trial is updated using the standard Elo rule
\[
R \leftarrow R + K'(S - E),
\]
where $E$ is the side’s expected score and $S$ is the observed score for that side. We use a base $K=32$ and scale it by the judge confidence $c\in[0,1]$ via $K' = K(0.5 + c)$, yielding larger updates for more decisive outcomes (i.e., $K'\in[16,48]$). All traits are initialized at 1500.

We maintain three Elo pools: (i) an \emph{overall} pool where traits are updated regardless of role, (ii) a \emph{prosecution-role} pool where traits are updated only when used by prosecution, and (iii) a \emph{defense-role} pool where traits are updated only when used by defense. Elo is computed separately for each experimental condition (mode/trait-count/round-depth/model), and we do not compare Elo values across different agent backends.

Because updates are applied to every trait that appears in a trial, Elo estimates reflect marginal performance of traits across many team contexts rather than a single fixed team composition.

\subsection{Trait Importance}

\begin{figure}
    \centering
    \includegraphics[width=1\linewidth]{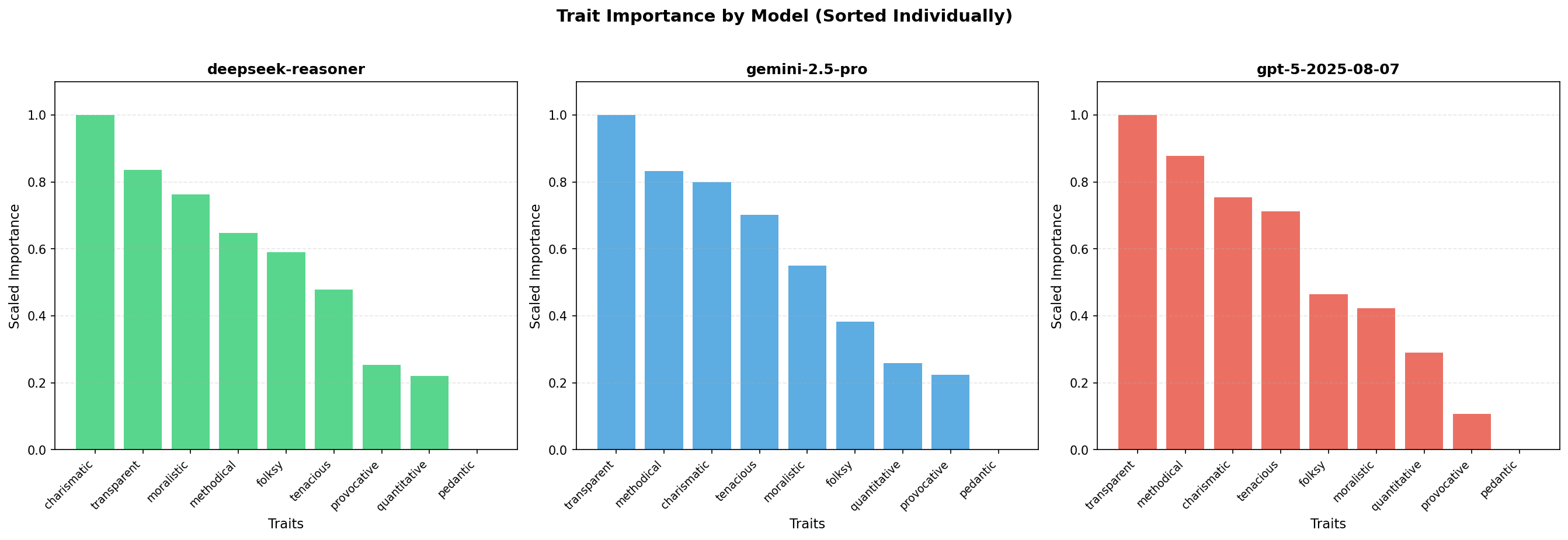}
    \caption{Normalized trait-importance scores derived from model-based rankings of three-trait combinations across DeepSeek-R1, Gemini-2.5-Pro, and GPT-5. Higher values indicate traits more frequently ranked as most important within winning configurations.}
    \label{fig:trait-importance}
\end{figure}

We have asked various models directly to rank all permutations of three traits of the pre-defined traits as outlined in Table \ref{tab:profiles}. Most important trait of three have been given a value of 2, second place got a value of 1 and least important trait got a value of 0. Those values have been summed up per trait and normalized -- we show this in Figure \ref{fig:trait-importance} for deepseek-reasoner, gemini-2.5-pro and gpt-5.

\subsection{Overall Performance Trends}

Across all models and configurations, heterogeneous teams achieve higher average Elo than single-agent or homogeneous baselines (Table~\ref{tab:aggregate}), supporting \textbf{H1}. In particular, two-trait teams consistently outperform single-trait configurations for both prosecution and defense.

Iterative interaction further improves outcomes: three-round debates yield the highest average Elo, while one- and two-round settings show weaker performance. These results indicate that moderate interaction depth enables adaptive strategy without inducing excessive repetition.

\begin{table}
\centering
\caption{Top Experiment Setups Ranked by Prosecution Elo (Best Trait per Setup)}
\label{tab:prosecution}
\begin{adjustbox}{width=\columnwidth}
\begin{tabular}{ccccccc}
\toprule
Rank & Mode & Traits & Rounds & Model & Top Elo & Best Trait \\
\midrule
1 & single & 2 & 3 & gemini-2.5-pro & 1789.4 & charismatic \\
2 & team   & 2 & 3 & gemini-2.5-pro & 1782.7 & charismatic \\
3 & team   & 1 & 2 & deepseek-reasoner & 1710.5 & quantitative \\
4 & single & 1 & 1 & deepseek-reasoner & 1706.0 & quantitative \\
5 & single & 1 & 3 & deepseek-reasoner & 1693.0 & transparent \\
6 & team   & 1 & 3 & deepseek-reasoner & 1691.8 & quantitative \\
7 & single & 1 & 2 & deepseek-reasoner & 1665.2 & charismatic \\
\bottomrule
\end{tabular}
\end{adjustbox}
\end{table}

\begin{table}[t]
\centering
\caption{Top Experiment Setups Ranked by Defense Elo (Best Trait per Setup)}
\label{tab:defense}
\begin{adjustbox}{width=\columnwidth}
\begin{tabular}{ccccccc}
\toprule
Rank & Mode & Traits & Rounds & Model & Top Elo & Best Trait \\
\midrule
1 & team   & 2 & 3 & gemini-2.5-pro & 1923.4 & quantitative \\
2 & single & 2 & 3 & gemini-2.5-pro & 1847.3 & quantitative \\
3 & team   & 1 & 3 & deepseek-reasoner & 1599.8 & quantitative \\
4 & team   & 1 & 2 & deepseek-reasoner & 1567.1 & tenacious \\
5 & single & 1 & 1 & deepseek-reasoner & 1547.3 & quantitative \\
6 & single & 1 & 2 & deepseek-reasoner & 1544.9 & quantitative \\
7 & single & 1 & 3 & deepseek-reasoner & 1530.6 & quantitative \\
\bottomrule
\end{tabular}
\end{adjustbox}
\end{table}

\begin{table}[t]
\centering
\caption{Top Experiment Setups Ranked by Overall Elo}
\label{tab:overall}
\begin{adjustbox}{width=\columnwidth}
\begin{tabular}{cccccc}
\toprule
Rank & Mode & Traits & Rounds & Model & Best Trait \\
\midrule
1 & team   & 2 & 3 & gemini-2.5-pro & quantitative \\
2 & single & 2 & 3 & gemini-2.5-pro & quantitative \\
3 & team   & 1 & 3 & deepseek-reasoner & quantitative \\
4 & single & 1 & 1 & deepseek-reasoner & quantitative \\
5 & team   & 1 & 2 & deepseek-reasoner & transparent \\
6 & single & 1 & 2 & deepseek-reasoner & quantitative \\
7 & single & 1 & 3 & deepseek-reasoner & methodical \\
\bottomrule
\end{tabular}
\end{adjustbox}
\end{table}

\begin{table}[t]
\centering
\caption{Best Performing Configurations Across Evaluation Dimensions}
\label{tab:bestconfigs}
\begin{adjustbox}{width=\columnwidth}
\begin{tabular}{llll}
\toprule
Category & Mode & Traits / Rounds / Model & Best Trait (Elo, Win Rate) \\
\midrule
Best Prosecution & single & 2 traits, 3 rounds, gemini-2.5-pro & charismatic (1789.4, 63.8\%) \\
Best Defense & team & 2 traits, 3 rounds, gemini-2.5-pro & quantitative (1923.4, 37.5\%) \\
Best Overall & team & 2 traits, 3 rounds, gemini-2.5-pro & quantitative (1893.5, --) \\
\bottomrule
\end{tabular}
\end{adjustbox}
\end{table}

\begin{table}[t]
\centering
\caption{Aggregate Elo by Experimental Dimension}
\label{tab:aggregate}
\begin{adjustbox}{width=\columnwidth}
\begin{tabular}{llll}
\toprule
Dimension & Category & Avg Prosecution Elo & Avg Defense Elo \\
\midrule
Mode & Single & 1713.4 & 1617.5 \\
Mode & Team & 1728.3 & 1696.8 \\
\midrule
Model & deepseek-reasoner & 1693.3 & 1558.0 \\
Model & gemini-2.5-pro & 1786.1 & 1885.3 \\
\midrule
Traits & 1 Trait & 1693.3 & 1558.0 \\
Traits & 2 Traits & 1786.1 & 1885.3 \\
\midrule
Rounds & 1 Round & 1706.0 & 1547.3 \\
Rounds & 2 Rounds & 1687.9 & 1556.0 \\
Rounds & 3 Rounds & 1739.2 & 1725.3 \\
\bottomrule
\end{tabular}
\end{adjustbox}
\end{table}

\subsection{Trait-Level Effects}

Trait frequency analysis over winning defense configurations reveals that \textbf{quantitative} and \textbf{charismatic} appear most consistently in top-ranked setups (Tables~\ref{tab:defense}, \ref{tab:overall}), supporting \textbf{H2}. Quantitative agents benefit from structured logical exposition, while charismatic agents contribute emotional salience and credibility.

Diplomat traits (\textit{transparent}, \textit{methodical}) show strong secondary effects, particularly when paired with either quantitative or charismatic teammates. By contrast, \textit{provocative} exhibits high variance: it is effective in select matchups but harmful when overused, suggesting it functions best as a situational rather than core strategy.

\subsection{Diversity Effects}

The strongest configurations combine traits from multiple archetypes. Common high-performing pairings include: \textbf{Rhetorician + Diplomat}: emotional appeal with structured explanation. \textbf{Technician + Gladiator}: logical rigor with persistent pressure.

Homogeneous teams display higher variance and greater sensitivity to case type, indicating narrower strategic coverage.

\subsection{Round Depth and Stability} \label{sec:round-depth-stability}

Outcome stability increases with additional rounds up to $N=3$. Re-evaluating identical cases shows a 23\% verdict reversal rate at $N=1$, decreasing to 8\% at $N=3$. Beyond five rounds, gains saturate while repetition increases, suggesting $N^* \approx 3$ as an effective operating point (RQ3).

\subsection{Model Sensitivity}

DeepSeek-R1 produces the most stable and high-confidence verdicts, consistent with its emphasis on explicit reasoning. Gemini~2.5~Pro exhibits greater sensitivity to rhetorically rich traits, benefiting Rhetorician-heavy teams. These differences highlight that optimal trait composition is partially model-dependent.

\subsection{Judge Reliability Analysis}\label{judge-reli}

To estimate judge consistency, we re-evaluate a random subset of trials with three independently seeded judge instances. Verdict agreement exceeds 85\%, suggesting moderate to high stability. These results indicate that observed performance differences are unlikely to be artifacts of stochastic judge variation. Nevertheless, LLM-based judges may introduce systematic biases, and human evaluation remains future work.

\subsection{Training a Trait Orchestrator} \label{sec:training-trait-orchestrator}

We investigate whether a learned trait-generation policy can outperform static, human-designed trait combinations by dynamically adapting defense team composition to the case context and opposing prosecution traits. To this end, we fine-tune a Qwen2.5-1.5B-Instruct model to act as a \emph{Trait Orchestrator} that generates three defense traits per trial.

\paragraph{Policy and Input.}
The orchestrator is prompted with structured case information (case name, summary, legal issues, and evidence) as well as the traits assigned to the opposing prosecution team. It is instructed to output exactly three unique trait names -- one per defense agent -- and is explicitly allowed to invent novel traits beyond the predefined taxonomy.

\paragraph{Training Procedure.}
We train the orchestrator using REINFORCE policy gradient. For each episode, the orchestrator proposes three traits for the defense team. The prosecution team samples three traits uniformly at random from the predefined nine-trait taxonomy. Both teams are instantiated using DeepSeek-R1 as the argumentative agent model and engage in a full simulated trial. The orchestrator receives a scalar reward based on the final verdict and judge confidence:
\[
R(v, c) =
\begin{cases}
+c & \text{if } v = \text{not guilty} \\
-c & \text{if } v = \text{guilty}
\end{cases}
\]
We apply LoRA fine-tuning (rank 16, $\alpha=32$) with 4-bit quantization and train for 500 episodes, selecting the best learning rate from $\{10^{-5}, 5\times10^{-5}, 10^{-4}\}$.

\paragraph{Learning Dynamics.}
Figure~3 shows cumulative reward over training, indicating a clear upward trend and stabilization after approximately 300 episodes. This suggests that the orchestrator learns a policy that reliably proposes defense trait combinations leading to more favorable outcomes than random initialization.

\paragraph{Performance Relative to Static Baselines.}
The learned orchestrator achieves higher average defense Elo than the best static two-trait and three-trait configurations discovered via exhaustive search. In particular, orchestrator-generated traits outperform static baselines in 62\% of matched evaluations, supporting H3.

\paragraph{Qualitative Behavior.}
Inspection of generated traits reveals that the orchestrator frequently produces semantically meaningful traits that resemble effective archetypes (e.g., \textit{evidence-weaver}, \textit{measured-skeptic}, \textit{calm-analyst}) as well as hybrid styles combining logical rigor and emotional grounding. Notably, the model does not simply reproduce the original nine traits, indicating discovery beyond the predefined taxonomy.

\begin{figure}[H]
    \centering
    \includegraphics[width=1\linewidth]{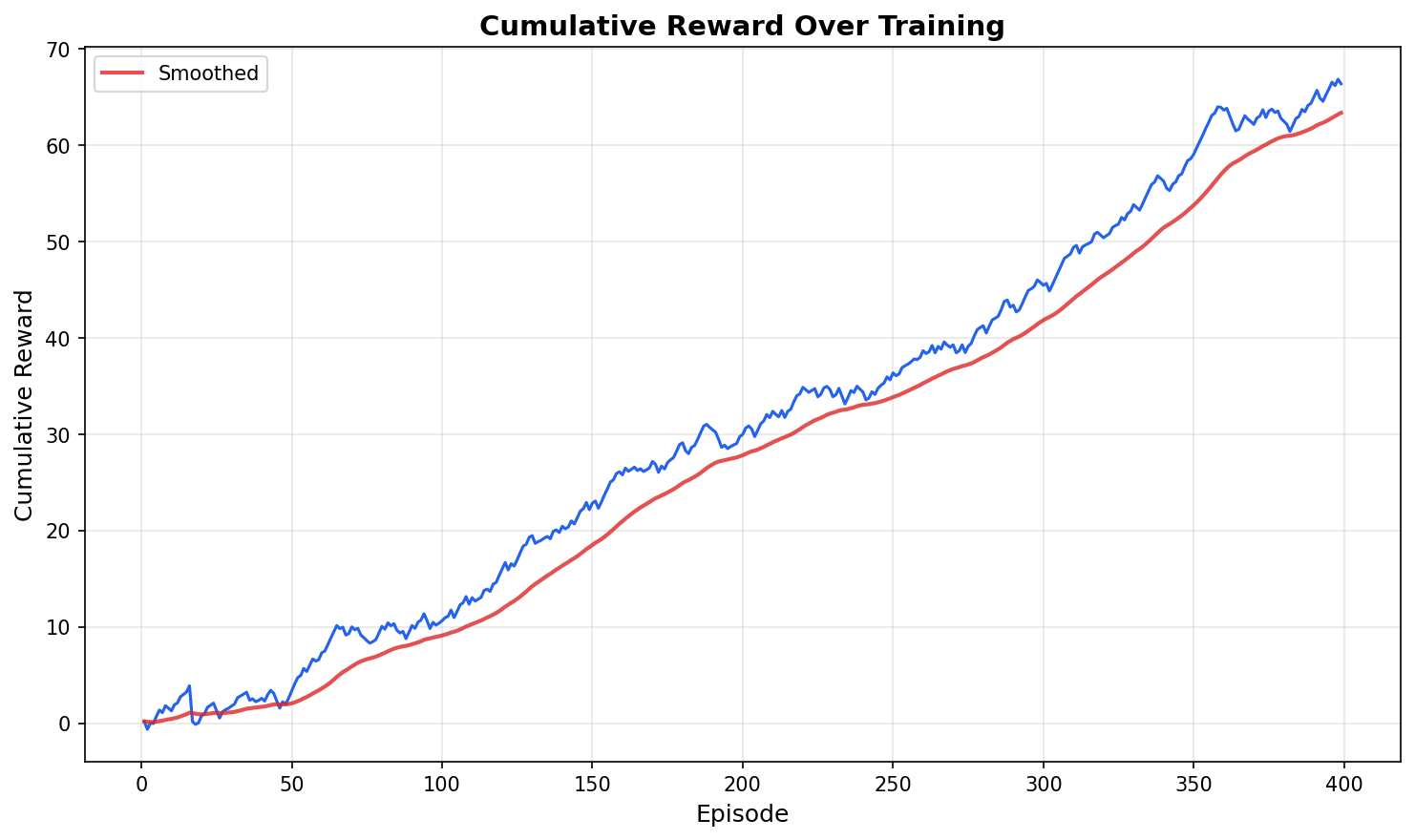}
    \caption{Cumulative reward of the RL-based Trait Orchestrator during training, showing convergence toward stable positive returns as trait generation improves.}
    \label{fig:cumulative-reward}
\end{figure}

\begin{figure}[H]
    \centering
    \includegraphics[width=1\linewidth]{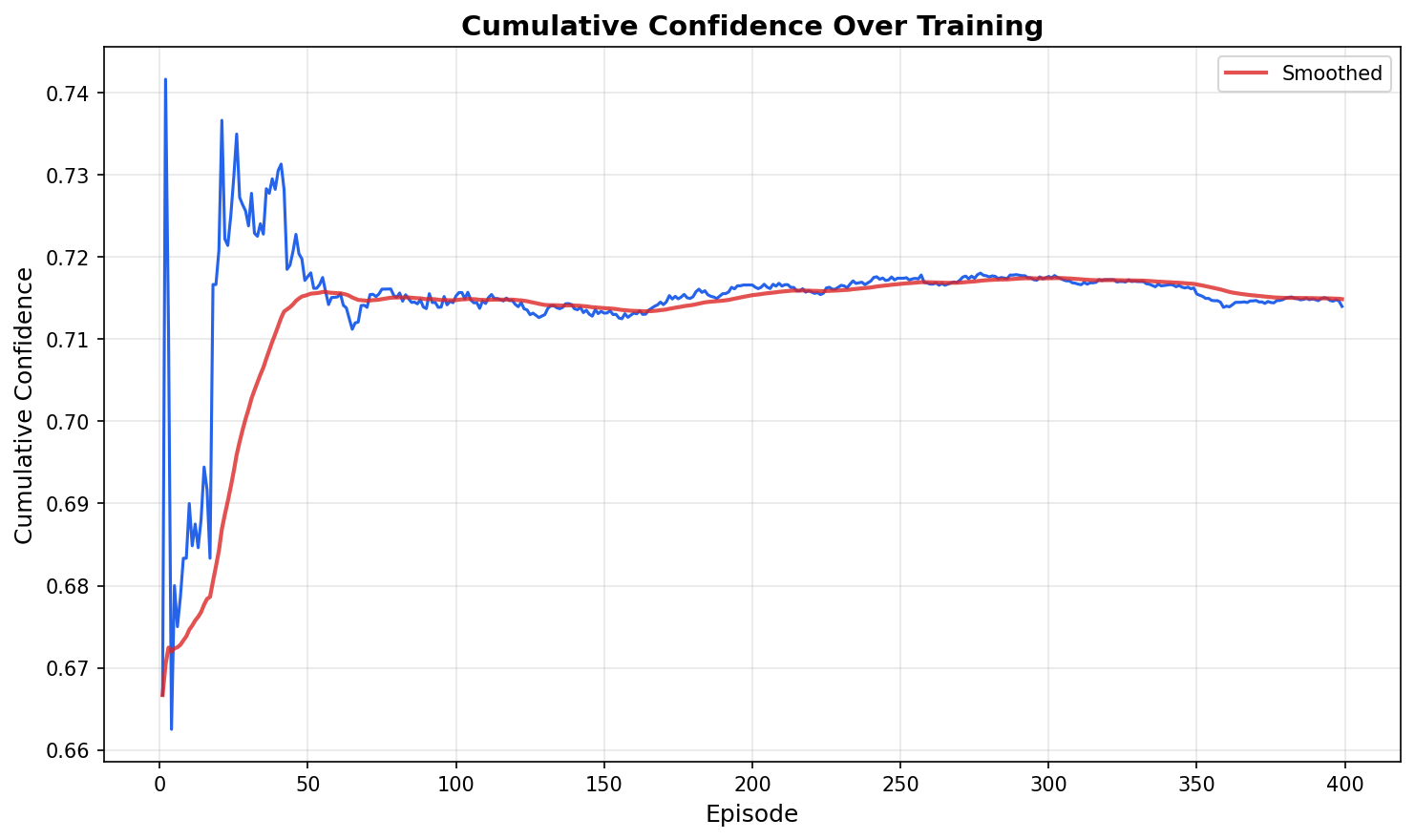}
    \caption{Cumulative judge confidence over training episodes for the RL-based Trait Orchestrator, illustrating increasing certainty in favorable defense outcomes.}
    \label{fig:cumulative-confidence}
\end{figure}

\begin{figure}[h]
    \centering
    \includegraphics[width=1\linewidth]{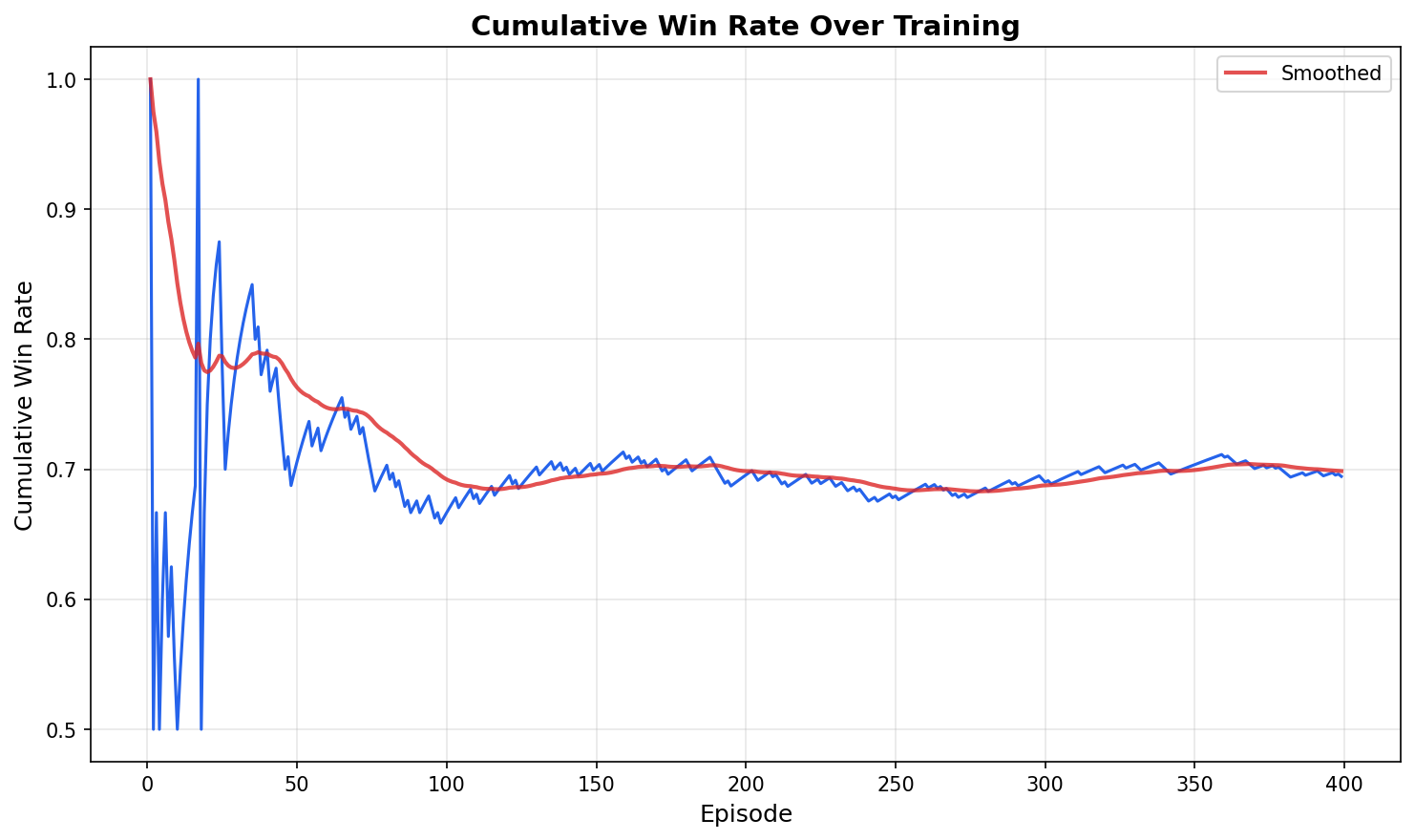}
    \caption{Cumulative defense win rate over training episodes for the RL-based Trait Orchestrator, demonstrating learning progress compared to early random performance.}
    \label{fig:cumulative-win-rate}
\end{figure}

\paragraph{Failure Modes.}
We observe two primary failure modes: (1) occasional generation of overly vague traits (e.g., \textit{smart}, \textit{strong}) that provide little behavioral guidance, and (2) collapse toward repeatedly generating a small subset of high-level trait concepts. These suggest the need for auxiliary diversity regularization and semantic filtering in future work.

\begin{table}[H]
\centering
\begin{tabular}{lcc}
\hline
Method & Avg Defense Elo & Win Rate (\%) \\
\hline
Best Static (2 Traits) & 1885.3 & 37.5 \\
Best Static (3 Traits) & 1869.1 & 36.2 \\
RL Orchestrator & \textbf{1912.4} & \textbf{41.1} \\
\hline
\end{tabular}
\caption{Performance comparison between best static trait configurations and the RL-based Trait Orchestrator.}
\end{table}

\section{Discussion: Towards Strategic Autonomy}

Our framework provides a blueprint for ``reducing the friction'' in the game-theoretic pipeline. By automating the translation of rich case summaries into a Strategic World Model (the trial flow), we demonstrate how LLM agents can handle the messiness of real-world negotiation while remaining amenable to strategic analysis.

\subsection{Emergent Strategic Behaviors}

We observed several emergent phenomena not explicitly programmed:
\begin{itemize}
    \item \textbf{Strategic Setup}: Provocative agents would destabilize opponents, creating openings for quantitative agents to present uncontested evidence.
    \item \textbf{Adaptive Framing}: Moralistic agents learned to reframe technical evidence in ethical terms when facing technician-heavy opposition.
    \item \textbf{Credibility Attacks}: Tenacious agents would persistently challenge opponent credibility across rounds, accumulating doubt.
\end{itemize}

\subsection{Limitations and Future Work}

Our current framework has several limitations that suggest directions for future research:

\textbf{Prompt Sensitivity}: Agent behavior depends on prompt wording; small changes in phrasing may alter trait expression and downstream outcomes.

\textbf{Synthetic Cases}: While our 10 cases span diverse legal domains, they lack the complexity of real proceedings. Future work should incorporate real case data with appropriate anonymization.

\textbf{Single Judge}: Real trials involve juries with heterogeneous preferences. Extending to multi-agent judicial panels would better model consensus formation.

\textbf{Static Traits}: Agents maintain fixed traits throughout trials. Adaptive trait switching mid-trial could enable more sophisticated strategies.

\textbf{Evaluation}: Our reliance on LLM judges introduces potential biases. Because judges share the same underlying model as advocates, they may implicitly favor linguistic patterns similar to those they generate; future work will explore cross-model evaluation and human-in-the-loop judging. Human evaluation studies would further strengthen validity claims.

\textbf{Synthetic Cases}: While our cases span diverse legal domains and are manually sanity-checked, they remain simplified abstractions of real proceedings. Future work should incorporate real or semi-synthetic case data with appropriate anonymization to evaluate external validity.

\subsection{Broader Applications}

This framework extends naturally to other adversarial discourse settings:
\begin{itemize}
    \item \textbf{Diplomatic Negotiations}: Modeling state actors with cultural and strategic profiles
    \item \textbf{Corporate Negotiations}: Simulating M\&A discussions with stakeholder traits
    \item \textbf{Consumer Advocacy}: Automated systems that argue for consumer interests against corporate policies
    \item \textbf{Educational Debate}: Training systems for competitive debate preparation
\end{itemize}

\section{Conclusion}

We introduced the Strategic Courtroom Framework, a multi-agent simulation environment for studying persuasion dynamics in adversarial legal proceedings using trait-conditioned LLM agents. By organizing agent behavior around an interpretable trait taxonomy and embedding agents within an iterative argumentation protocol, our framework enables systematic analysis of how personality, team composition, and interaction depth shape persuasive success.

Across more than 7{,}000 simulated trials, we showed that heterogeneous teams with complementary traits consistently outperform homogeneous configurations, and that moderate interaction depth yields more stable and reliable outcomes. Moreover, our reinforcement-learning-based Trait Orchestrator demonstrates that adaptive trait generation can surpass static, human-designed strategies, suggesting a path toward autonomous strategic agents that learn how to argue effectively rather than merely what to argue.

More broadly, this work advances a shift from viewing language as an unstructured medium toward treating it as a first-class strategic action space. As LLMs continue to improve, such environments provide a principled testbed for exploring negotiation, conflict resolution, and collective decision-making in domains where persuasion, not payoff matrices alone, determines outcomes. We release our code and experimental artifacts to support future research in this direction.

\balance

\newpage

\begin{acks}
We thank the anonymous reviewers for their constructive feedback.
\end{acks}

\bibliographystyle{ACM-Reference-Format} 
\bibliography{references}

\appendix
\onecolumn

\section*{Appendix}

\section{Trait Definitions}
\label{app:traits}

Table~\ref{tab:full_traits} provides complete definitions for all 9 character traits used in our experiments.

\begin{table}[h]
\caption{Complete Trait Definitions}
\label{tab:full_traits}
\centering
\begin{tabularx}{\textwidth}{llX}
\toprule
\textbf{Trait} & \textbf{Archetype} & \textbf{Full Description} \\
\midrule
Charismatic & Rhetorician & Appeals to the audience's emotions and rapport to sway judgment beyond mere facts. \\
Folksy & Rhetorician & The ``Mean'' of friendliness; appearing as a peer to the jury to foster trust. \\
Moralistic & Rhetorician & Frames the case through the lens of ``The Good,'' focusing on ultimate justice. \\
\midrule
Pedantic & Technician & Extreme focus on the ``letter'' of the law, often at the expense of the ``spirit'' or equity. \\
Quantitative & Technician & Relies on logical demonstration and hard data to prove a point (Syllogistic reasoning). \\
\midrule
Tenacious & Gladiator & The virtue of persisting in a difficult course of action despite legal or social pressure. \\
Provocative & Gladiator & Deliberately stirring up anger or conflict to gain a tactical advantage. \\
\midrule
Transparent & Diplomat & The ``Mean'' between self-deprecation and boastfulness; presenting the case exactly as it is. \\
Methodical & Diplomat & Using practical wisdom to guide the jury through a complex sequence of cause and effect. \\
\bottomrule
\end{tabularx}
\end{table}

\section{Case Corpus}
\label{app:cases}

Our synthetic case corpus covers 10 legal scenarios:

\begin{enumerate}

\item \textbf{State v. John Doe} \\
\textit{Summary:} Assault charge after an altercation at work. Defendant claims self-defense. \\
\textit{Evidence:}
\begin{itemize}
    \item Witness testimony from co-workers
    \item Security camera footage
    \item Medical report of victim’s injuries
\end{itemize}
\textit{Legal Issues:} Self-defense, Assault \\
\textit{Roles:} Prosecution, Defense, Judge, Evidence Analyzer

\item \textbf{Greenfield Corp. v. Alex Cruz} \\
\textit{Summary:} Greenfield alleges failure to deliver contracted software. Cruz claims requirements were incomplete. \\
\textit{Evidence:}
\begin{itemize}
    \item Contract agreement
    \item Emails between parties
    \item Project timeline and delivery logs
\end{itemize}
\textit{Legal Issues:} Breach of contract, Contractual obligations \\
\textit{Roles:} Plaintiff, Defendant, Judge

\item \textbf{State v. Rita Holmes} \\
\textit{Summary:} Charged with shoplifting. Holmes claims the incident was accidental. \\
\textit{Evidence:}
\begin{itemize}
    \item Store CCTV footage
    \item Receipt showing unpaid items
    \item Store clerk witness statement
\end{itemize}
\textit{Legal Issues:} Theft, Intent to steal \\
\textit{Roles:} Prosecution, Defense, Judge

\item \textbf{Smith v. Rodriguez} \\
\textit{Summary:} Smith alleges Rodriguez spread false rumors. Rodriguez argues statements were opinion. \\
\textit{Evidence:}
\begin{itemize}
    \item Social media posts
    \item Witness testimony
    \item Evidence of lost job opportunities
\end{itemize}
\textit{Legal Issues:} Defamation, Freedom of speech, Reputation damages \\
\textit{Roles:} Plaintiff, Defendant, Judge

\item \textbf{Anderson v. Larson Realty} \\
\textit{Summary:} Tenant claims unlawful eviction despite timely rent payments. Landlord alleges lease violations. \\
\textit{Evidence:}
\begin{itemize}
    \item Lease agreement
    \item Photos of property damage
    \item Rent payment records
\end{itemize}
\textit{Legal Issues:} Tenant rights, Lease compliance, Unlawful eviction \\
\textit{Roles:} Plaintiff, Defendant, Judge

\item \textbf{People v. Terry Nguyen} \\
\textit{Summary:} Charged with DUI. Nguyen claims sobriety test was improperly administered. \\
\textit{Evidence:}
\begin{itemize}
    \item Police report
    \item Sobriety test results
    \item Passenger testimony
\end{itemize}
\textit{Legal Issues:} DUI, Sobriety test procedure, Improper evidence administration \\
\textit{Roles:} Prosecution, Defense, Judge

\item \textbf{Jones v. BrightView Hospital} \\
\textit{Summary:} Surgical error allegedly caused permanent nerve damage. Hospital cites informed consent. \\
\textit{Evidence:}
\begin{itemize}
    \item Medical records
    \item Surgeon’s notes
    \item Signed consent form
\end{itemize}
\textit{Legal Issues:} Medical malpractice, Informed consent, Standard of care \\
\textit{Roles:} Plaintiff, Defendant, Judge

\item \textbf{City v. Ben Foster} \\
\textit{Summary:} Repeated noise violations reported by neighbors. Foster disputes volume level. \\
\textit{Evidence:}
\begin{itemize}
    \item Noise complaint records
    \item Neighbor audio recordings
    \item Foster’s decibel recordings
\end{itemize}
\textit{Legal Issues:} Noise ordinance violation, Community disturbance \\
\textit{Roles:} Prosecution, Defense, Judge

\item \textbf{Emily Park v. Phoenix Corp.} \\
\textit{Summary:} Park alleges gender discrimination in promotion decision. Company cites superior qualifications of selected candidate. \\
\textit{Evidence:}
\begin{itemize}
    \item Performance records
    \item Management emails
    \item Candidate qualifications
\end{itemize}
\textit{Legal Issues:} Gender discrimination, Employment law, Promotion criteria \\
\textit{Roles:} Plaintiff, Defendant, Judge

\item \textbf{Taylor v. Rustic Restaurants} \\
\textit{Summary:} Slip-and-fall injury allegedly caused by wet floor. Restaurant claims warning sign was posted. \\
\textit{Evidence:}
\begin{itemize}
    \item Medical records
    \item CCTV footage
    \item Employee testimony
\end{itemize}
\textit{Legal Issues:} Personal injury, Negligence, Adequate warnings \\
\textit{Roles:} Plaintiff, Defendant, Judge

\end{enumerate}

Each case includes structured evidence items and legal issues to ground agent argumentation.

\newpage

\section{Sample Trial Transcript}
\label{app:transcript}

Below is an abbreviated transcript from \textit{State v. John Doe} with prosecution traits (charismatic, folksy, moralistic) vs. defense traits (charismatic, folksy, pedantic):

\textbf{Prosecution Opening:}
\begin{quote}
\textit{``Good morning. Now, we all know that a workplace ought to be a place of safety. It's where we earn our bread, share a coffee, and trust one another. What the evidence will show is simple. John Doe didn't just have a disagreement; he made a choice. After words were exchanged, he followed his co-worker and threw the first punch. That's not self-defense; that's an attack.''}
\end{quote}

\textbf{Defense Opening:}
\begin{quote}
\textit{``Well now, ladies and gentlemen of the jury, what we have here is a story with two sides. My client, John Doe, is a hard-working fella who found himself in a spot no one wants to be in: backed into a corner. John acted for one reason, and one reason only: to protect himself when that heated argument crossed a line into a genuine threat.''}
\end{quote}

\textbf{Verdict:} Not Guilty (Confidence: 0.65)

\newpage

\section{Trial Experiments}

All appendix figures follow a consistent structure: confidence profiles, Elo rankings, Elo--win-rate relationships, and role-based effectiveness comparisons for each experimental setting.

\subsection{Single Agent - 1 Trait - 1 Round - DeepSeek-R1}

\begin{figure}[H]
    \centering
    \includegraphics[width=0.63\linewidth]{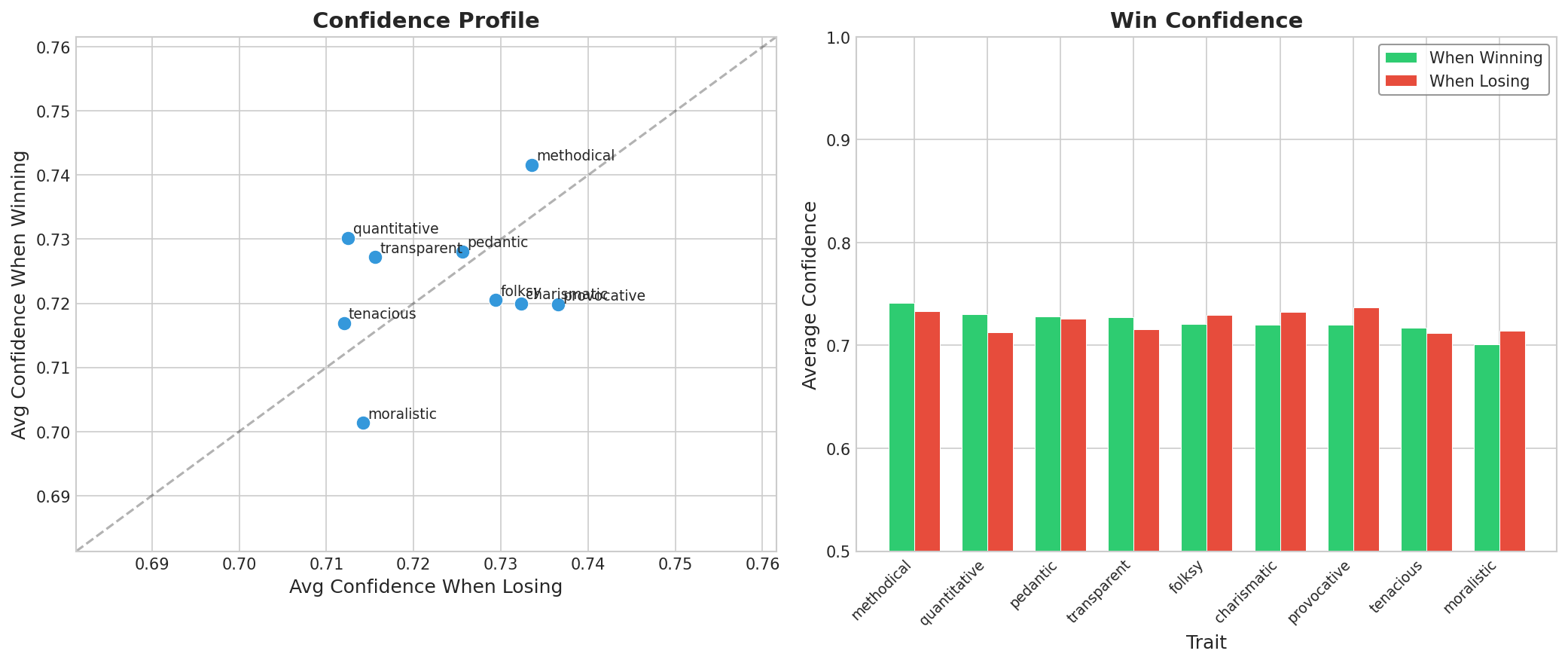}
    \caption{Trait confidence analysis for Single Agent, 1 Trait, 1 Round using DeepSeek-R1. Left: Average judge confidence when a trait wins versus loses. Right: Mean confidence across traits for winning and losing cases.}
\end{figure}

\begin{figure}[H]
    \centering
    \includegraphics[width=0.51\linewidth]{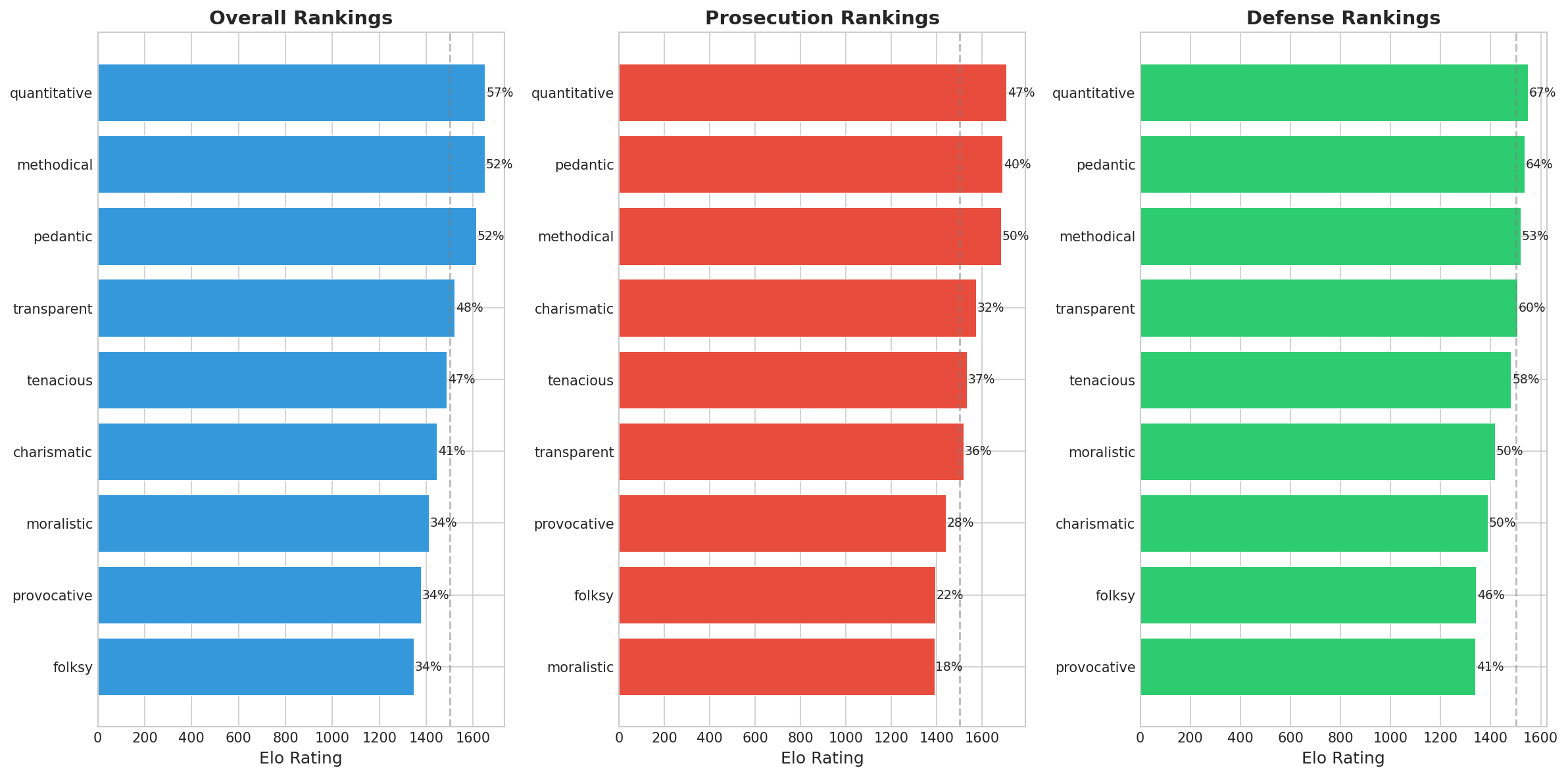}
    \caption{Elo rankings of traits under Single Agent, 1 Trait, 1 Round using DeepSeek-R1, shown for overall performance, prosecution role, and defense role.}
\end{figure}

\begin{figure}[H]
\centering
\begin{minipage}{0.3\linewidth}
\centering
\includegraphics[width=\linewidth]{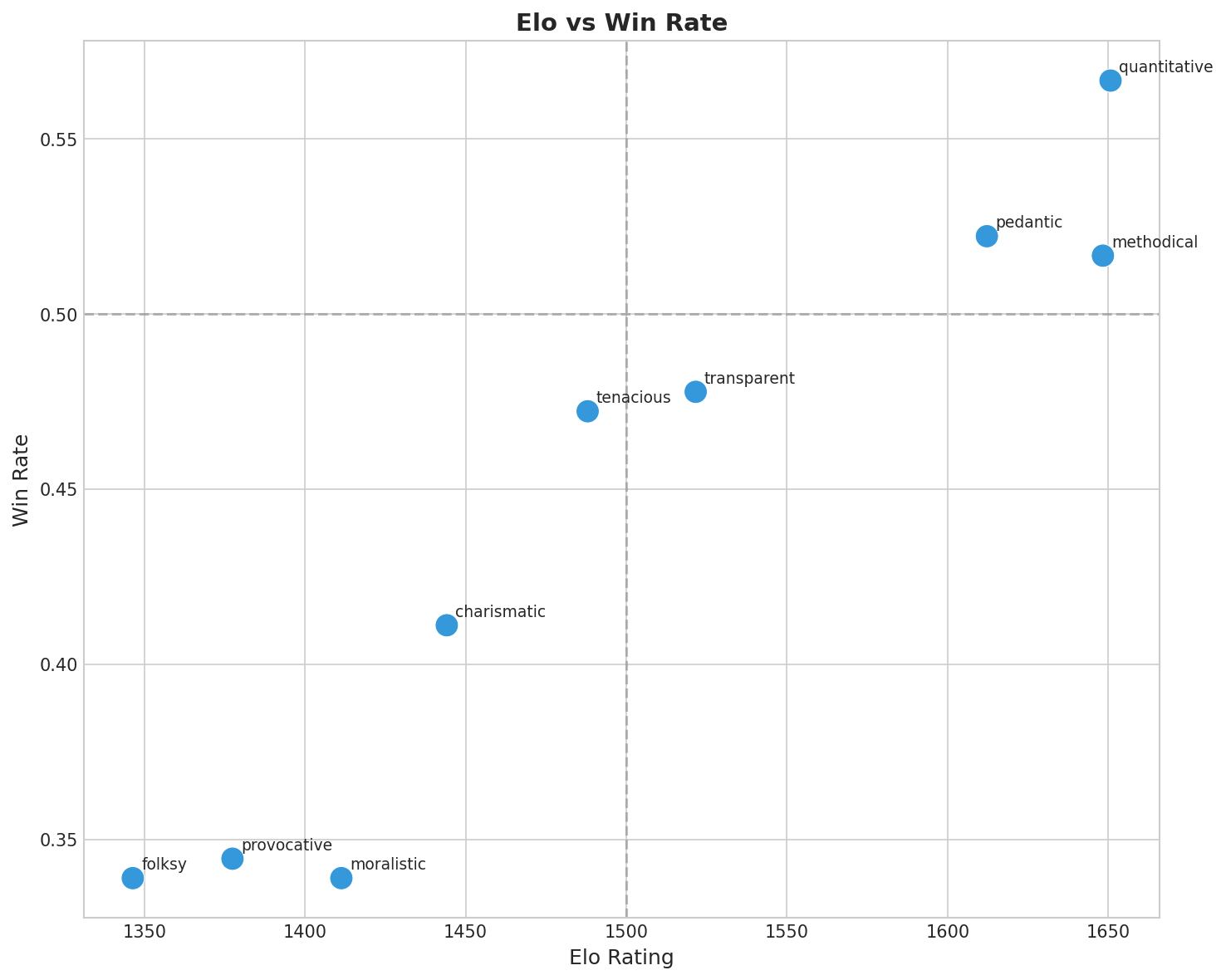}
\caption{Relationship between trait Elo rating and win rate under Single Agent, 1 Trait, 1 Round using DeepSeek-R1. Each point represents a trait.}
\end{minipage}
\hspace{0.1cm}
\begin{minipage}{0.3\linewidth}
\centering
\includegraphics[width=\linewidth]{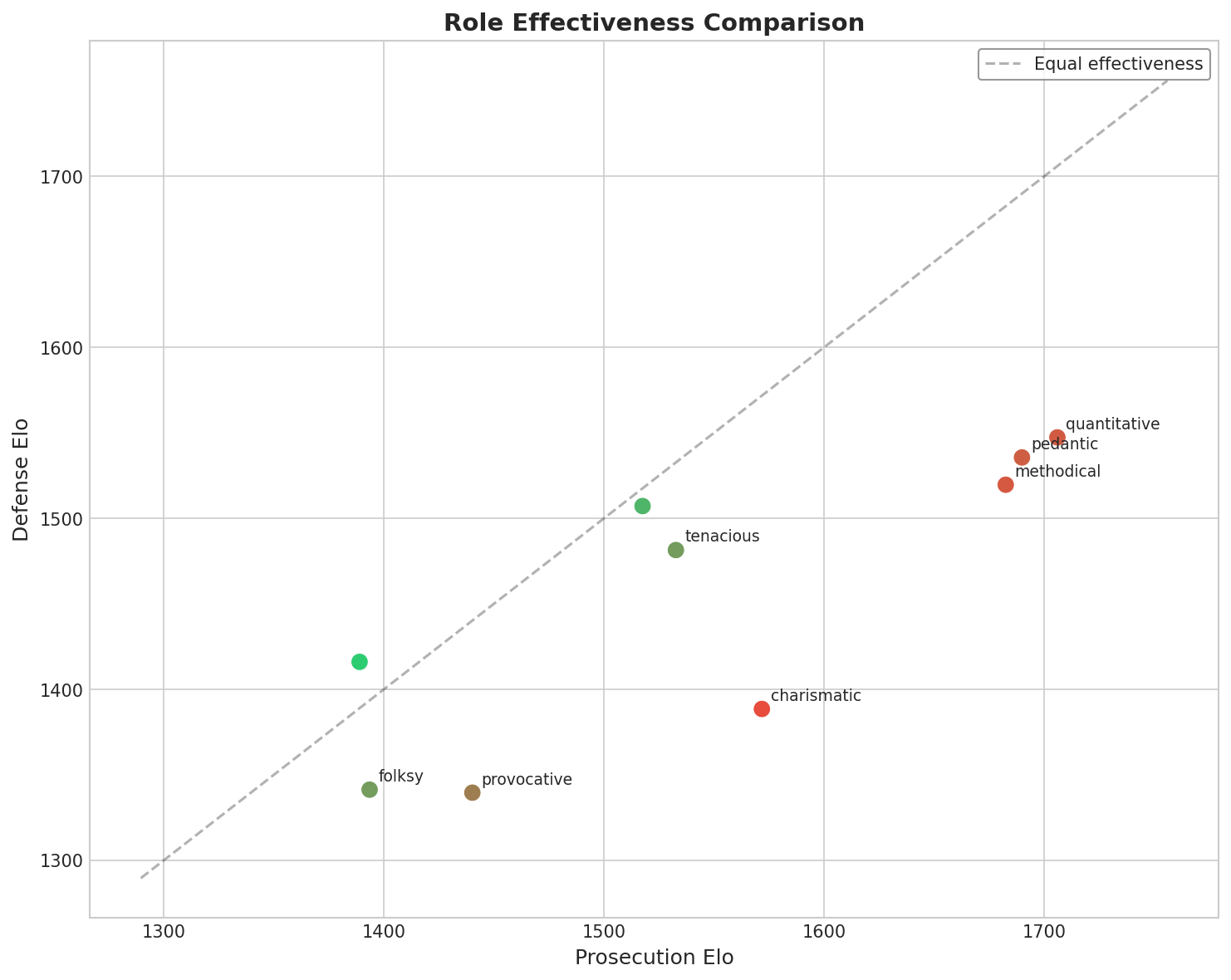}
\caption{Comparison of trait effectiveness across agent roles under Single Agent, 1 Trait, 1 Round using DeepSeek-R1, showing prosecution versus defense Elo for each trait.}
\end{minipage}
\end{figure}

\newpage

\subsection{Single Agent - 1 Trait - 2 Rounds - DeepSeek-R1}

\begin{figure}[H]
\centering
\includegraphics[width=0.63\linewidth]{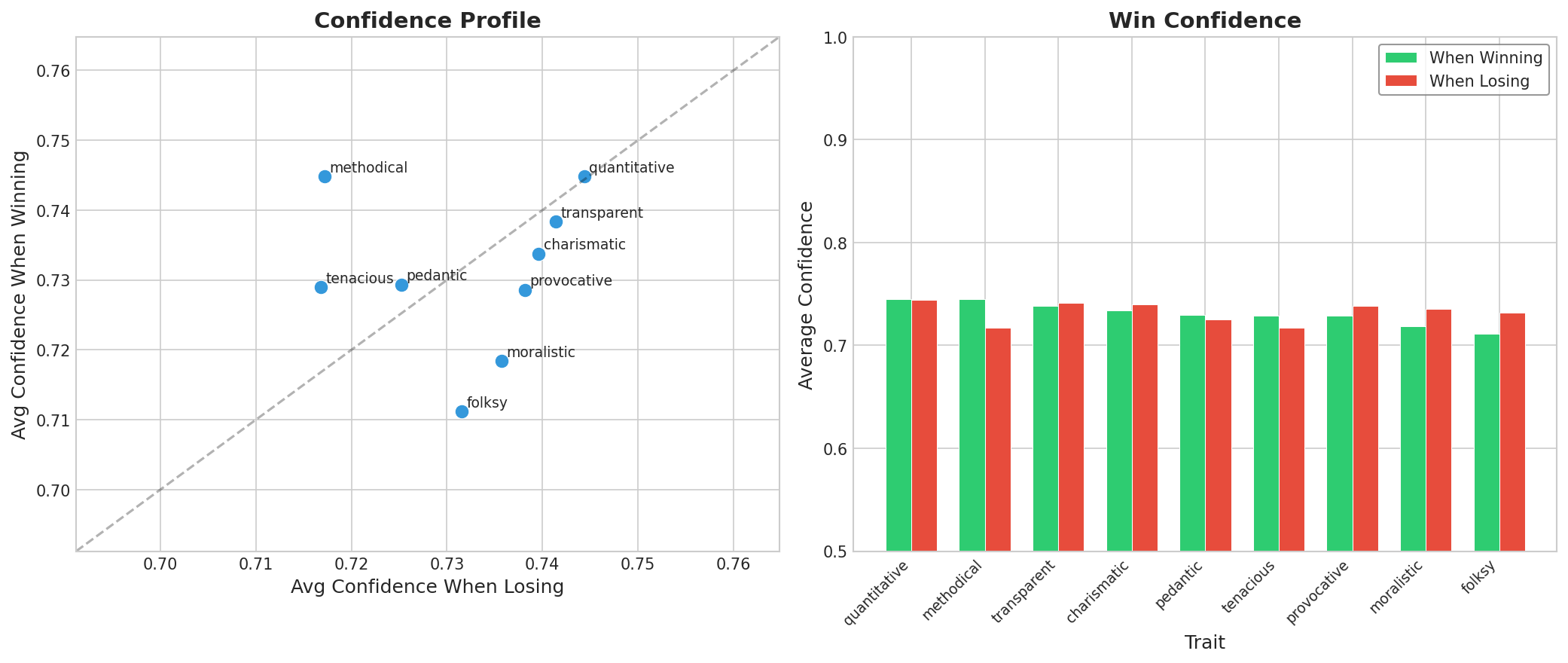}
\caption{Trait confidence analysis for Single Agent, 1 Trait, 2 Rounds using DeepSeek-R1. Left: Average judge confidence when a trait wins versus loses. Right: Mean confidence across traits for winning and losing cases.}
\end{figure}

\begin{figure}[H]
\centering
\includegraphics[width=0.51\linewidth]{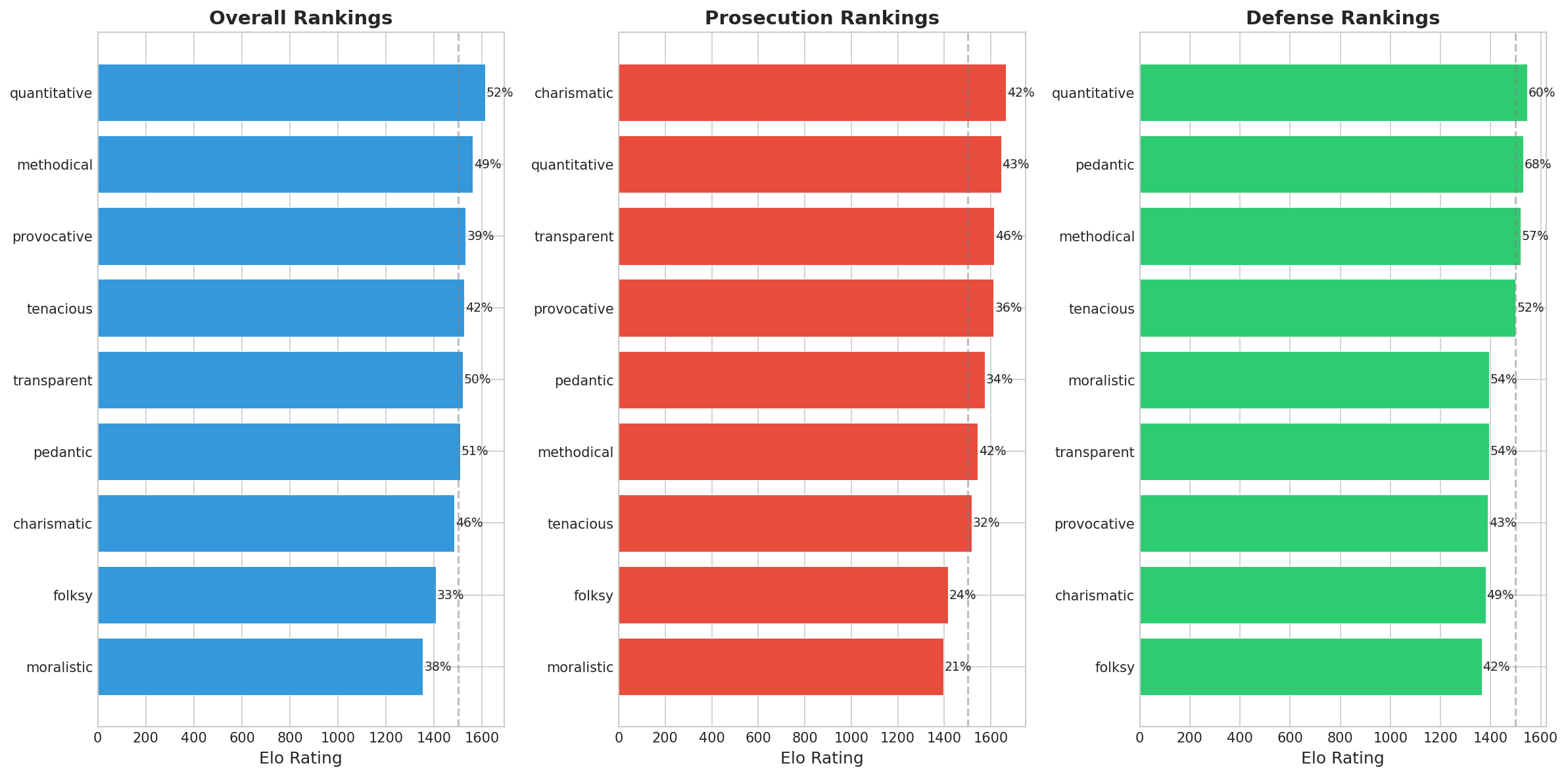}
\caption{Elo rankings of traits under Single Agent, 1 Trait, 2 Rounds using DeepSeek-R1, shown for overall performance, prosecution role, and defense role.}
\end{figure}

\begin{figure}[H]
\centering
\begin{minipage}{0.3\linewidth}
\centering
\includegraphics[width=\linewidth]{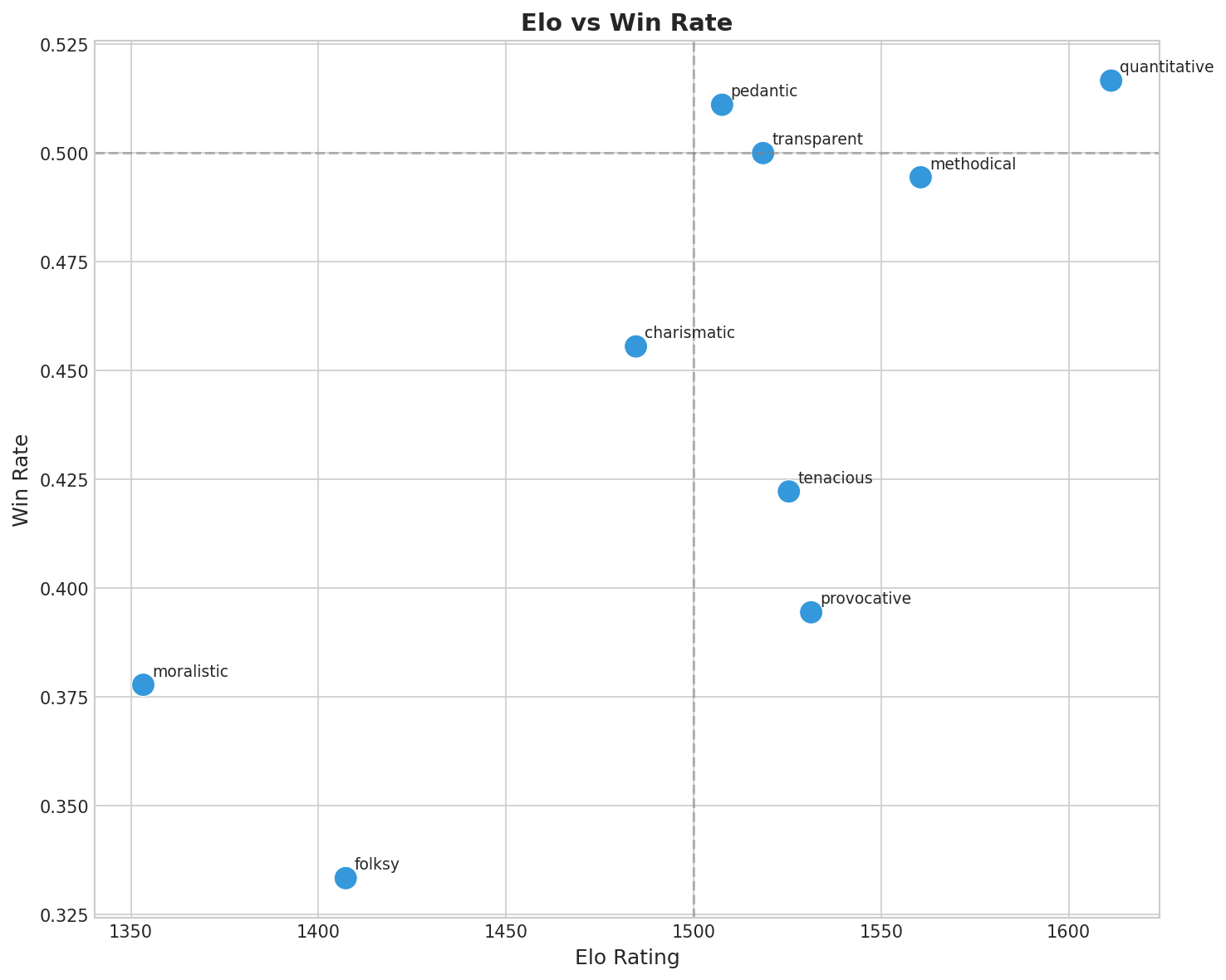}
\caption{Relationship between trait Elo rating and win rate under Single Agent, 1 Trait, 2 Rounds using DeepSeek-R1. Each point represents a trait.}
\end{minipage}
\hspace{0.1cm}
\begin{minipage}{0.3\linewidth}
\centering
\includegraphics[width=\linewidth]{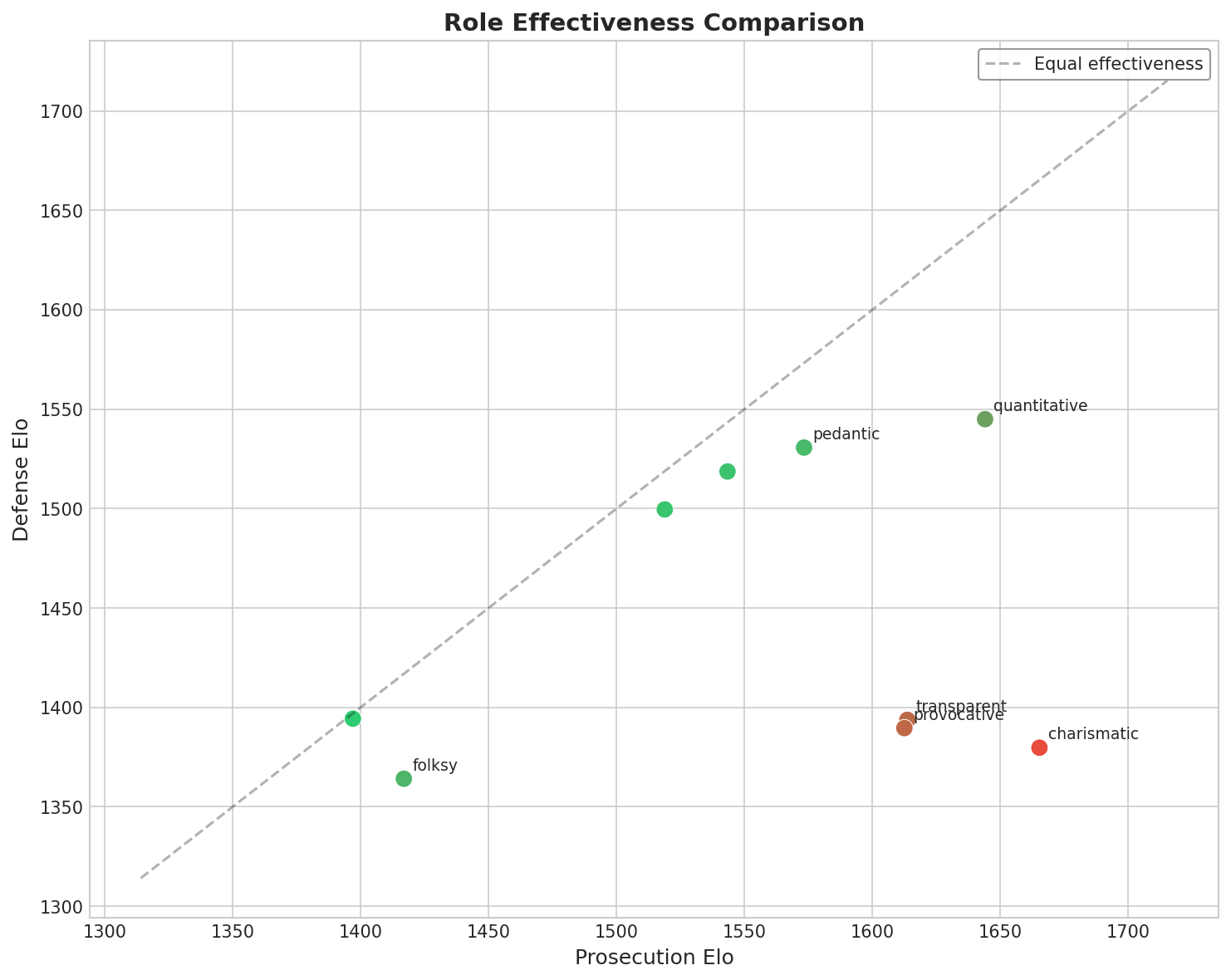}
\caption{Comparison of trait effectiveness across agent roles under Single Agent, 1 Trait, 2 Rounds using DeepSeek-R1, showing prosecution versus defense Elo for each trait.}
\end{minipage}
\end{figure}

\newpage

\subsection{Single Agent - 1 Trait - 3 Rounds - DeepSeek-R1}

\begin{figure}[H]
\centering
\includegraphics[width=0.63\linewidth]{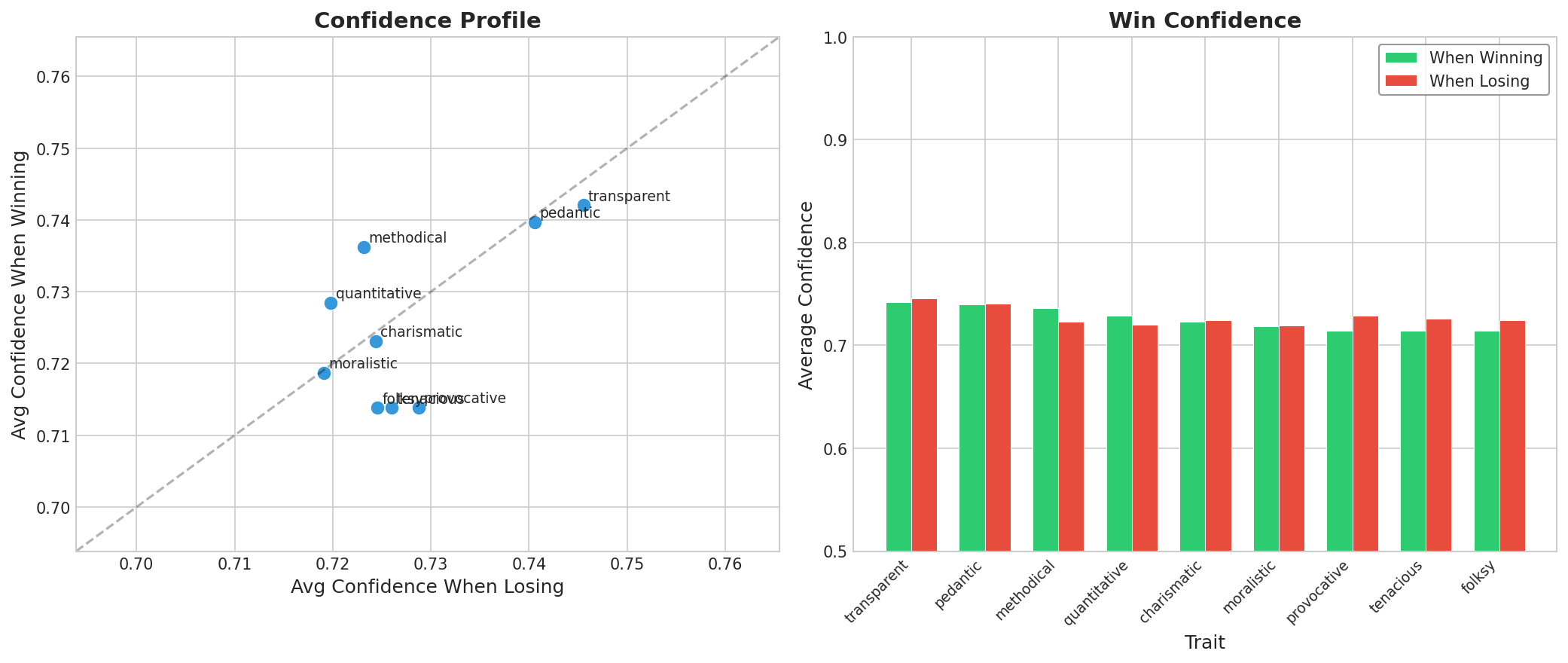}
\caption{Trait confidence analysis for Single Agent, 1 Trait, 3 Rounds using DeepSeek-R1. Left: Average judge confidence when a trait wins versus loses. Right: Mean confidence across traits for winning and losing cases.}
\end{figure}

\begin{figure}[H]
\centering
\includegraphics[width=0.51\linewidth]{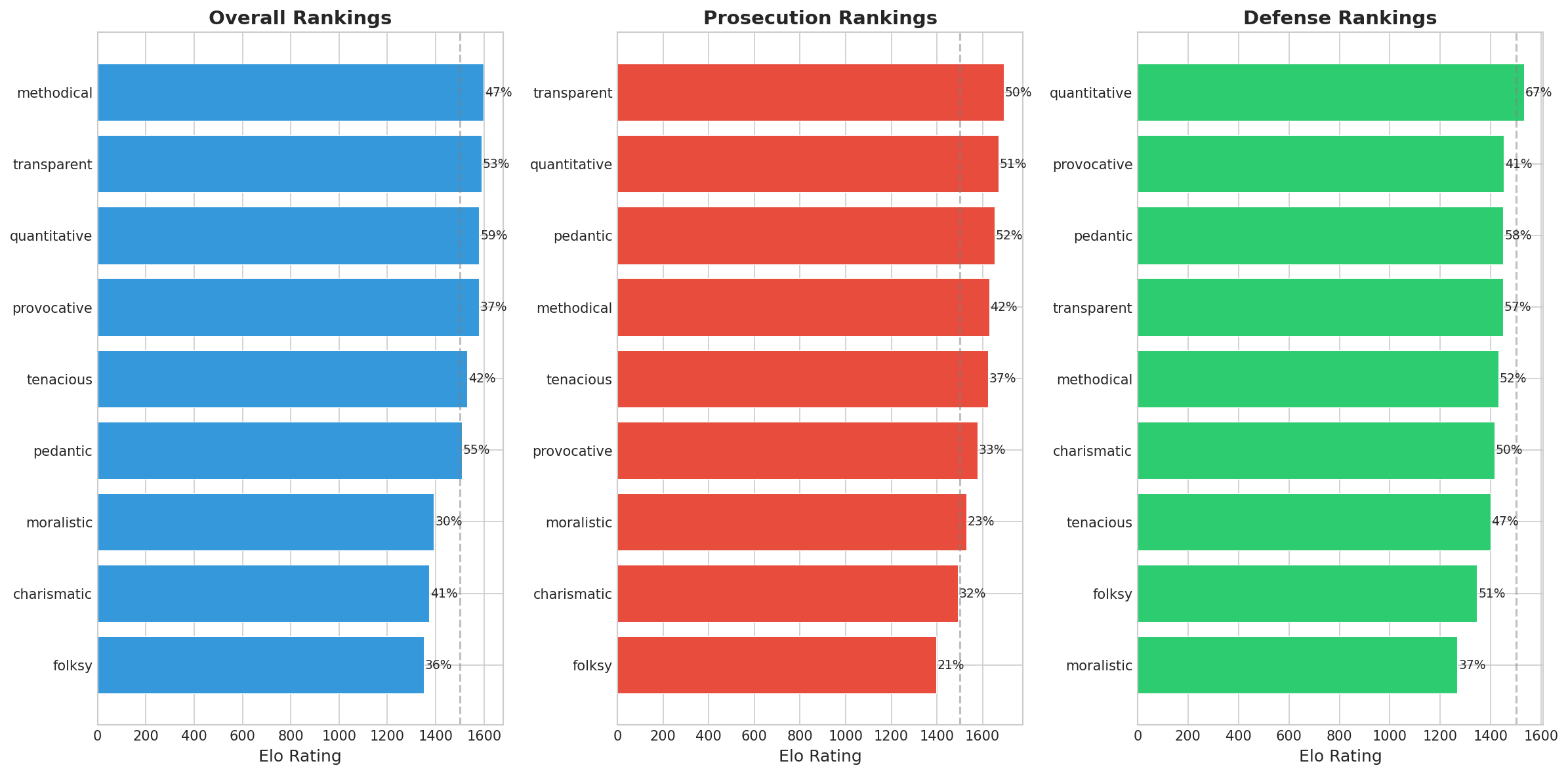}
\caption{Elo rankings of traits under Single Agent, 1 Trait, 3 Rounds using DeepSeek-R1, shown for overall performance, prosecution role, and defense role.}
\end{figure}

\begin{figure}[H]
\centering
\begin{minipage}{0.3\linewidth}
\centering
\includegraphics[width=\linewidth]{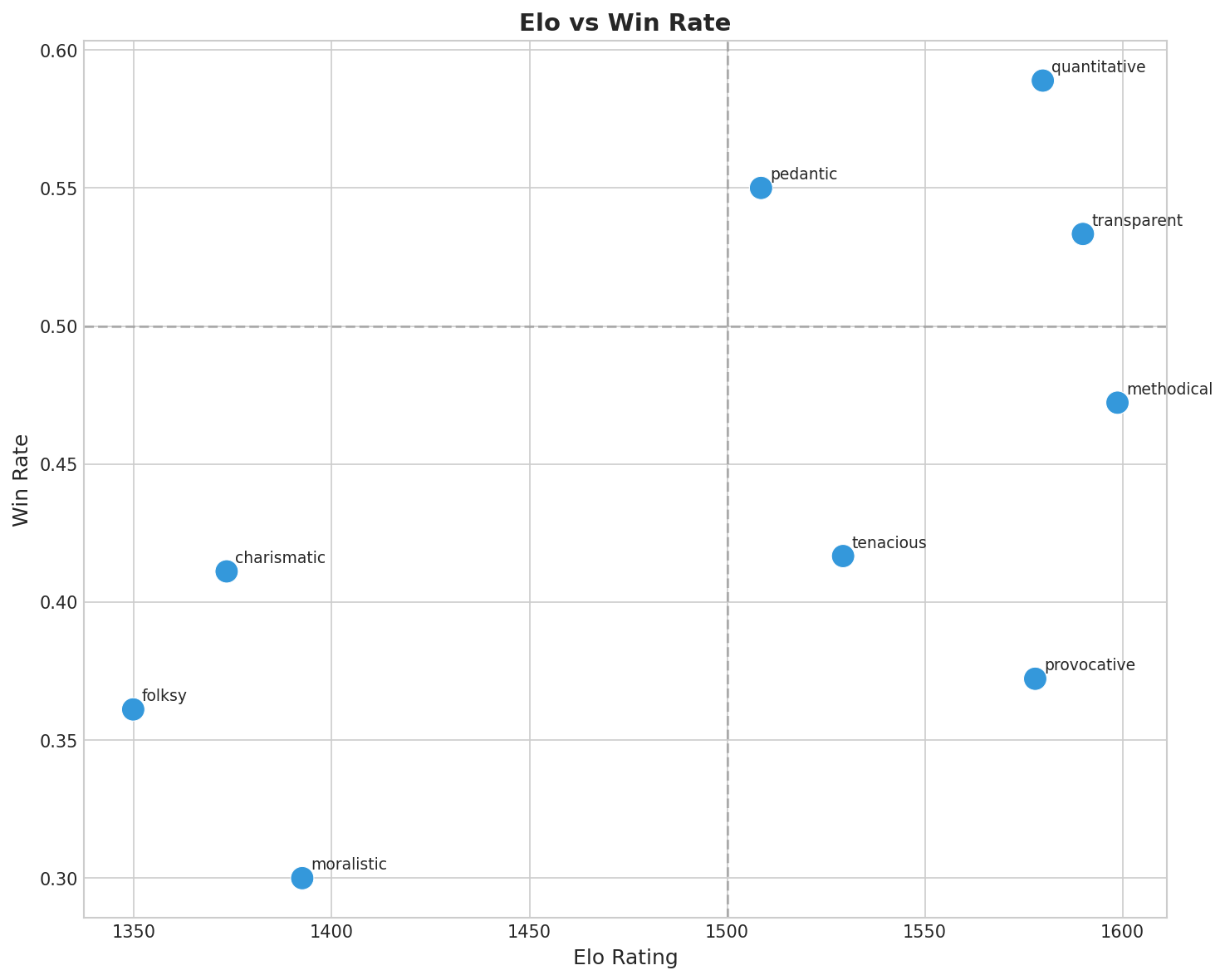}
\caption{Relationship between trait Elo rating and win rate under Single Agent, 1 Trait, 3 Rounds using DeepSeek-R1. Each point represents a trait.}
\end{minipage}
\hspace{0.1cm}
\begin{minipage}{0.3\linewidth}
\centering
\includegraphics[width=\linewidth]{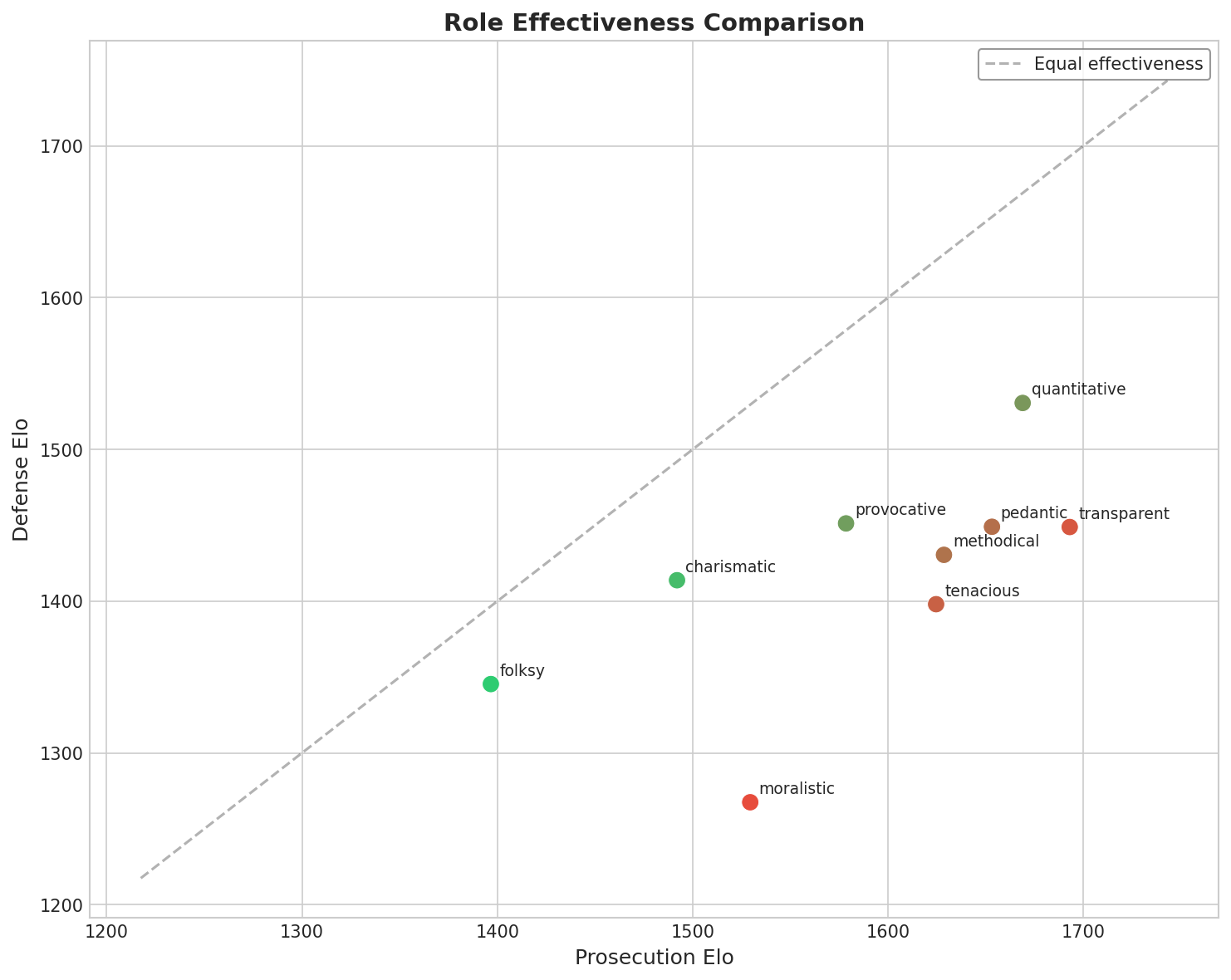}
\caption{Comparison of trait effectiveness across agent roles under Single Agent, 1 Trait, 3 Rounds using DeepSeek-R1, showing prosecution versus defense Elo for each trait.}
\end{minipage}
\end{figure}

\newpage

\subsection{Single Agent - 2 Traits - 3 Rounds - Gemini-2.5-Pro}

\begin{figure}[H]
\centering
\includegraphics[width=0.63\linewidth]{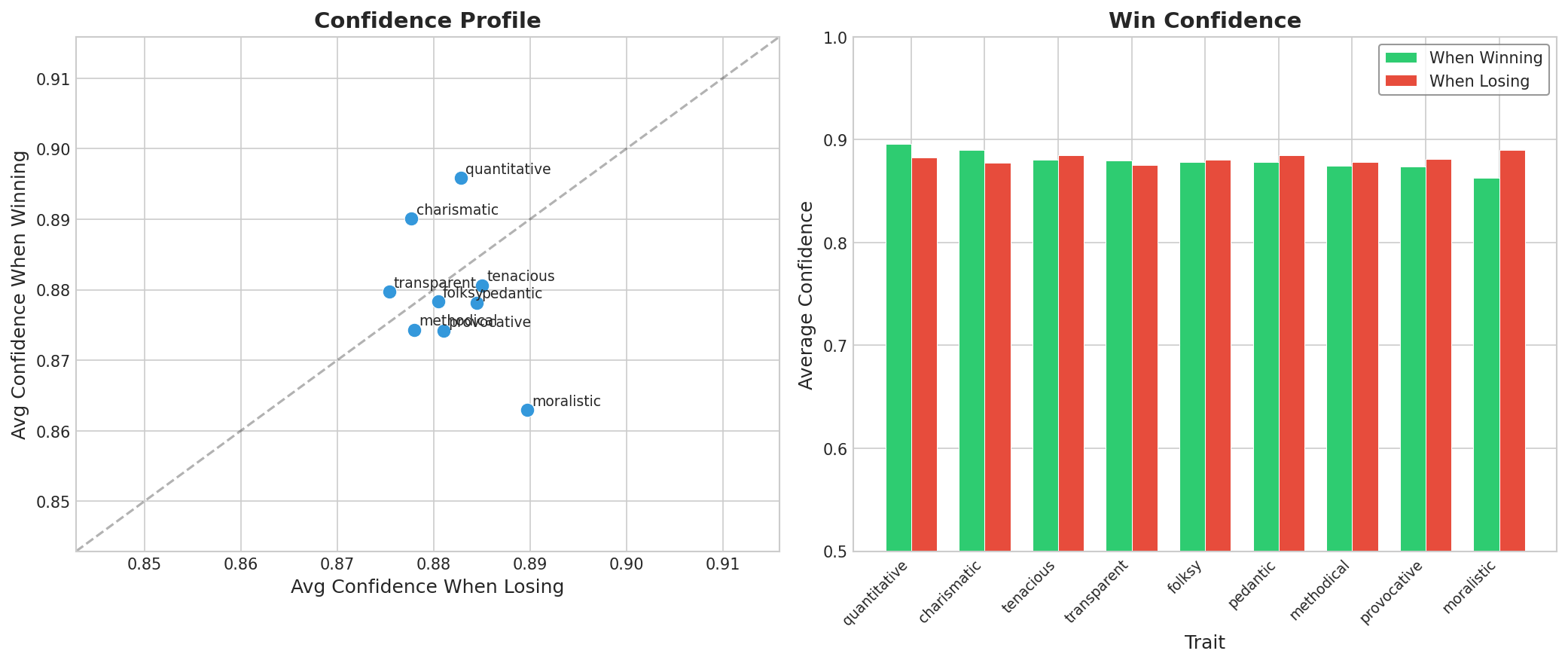}
\caption{Trait confidence analysis for Single Agent, 2 Traits, 3 Rounds using Gemini-2.5-Pro. Left: Average judge confidence when a trait wins versus loses. Right: Mean confidence across traits for winning and losing cases.}
\end{figure}

\begin{figure}[H]
\centering
\includegraphics[width=0.51\linewidth]{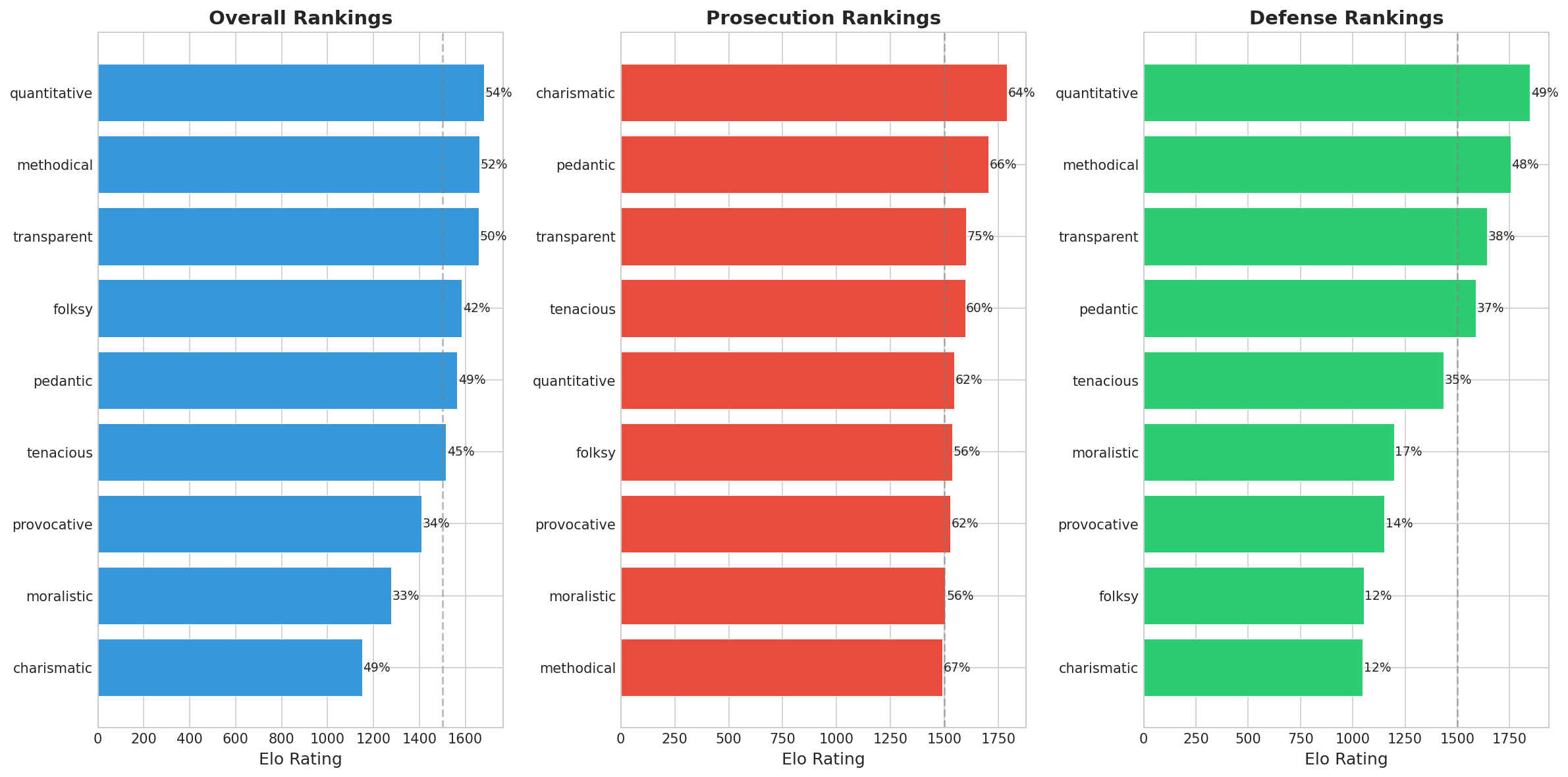}
\caption{Elo rankings of traits under Single Agent, 2 Traits, 3 Rounds using Gemini-2.5-Pro, shown for overall performance, prosecution role, and defense role.}
\end{figure}

\begin{figure}[H]
\centering
\begin{minipage}{0.3\linewidth}
\centering
\includegraphics[width=\linewidth]{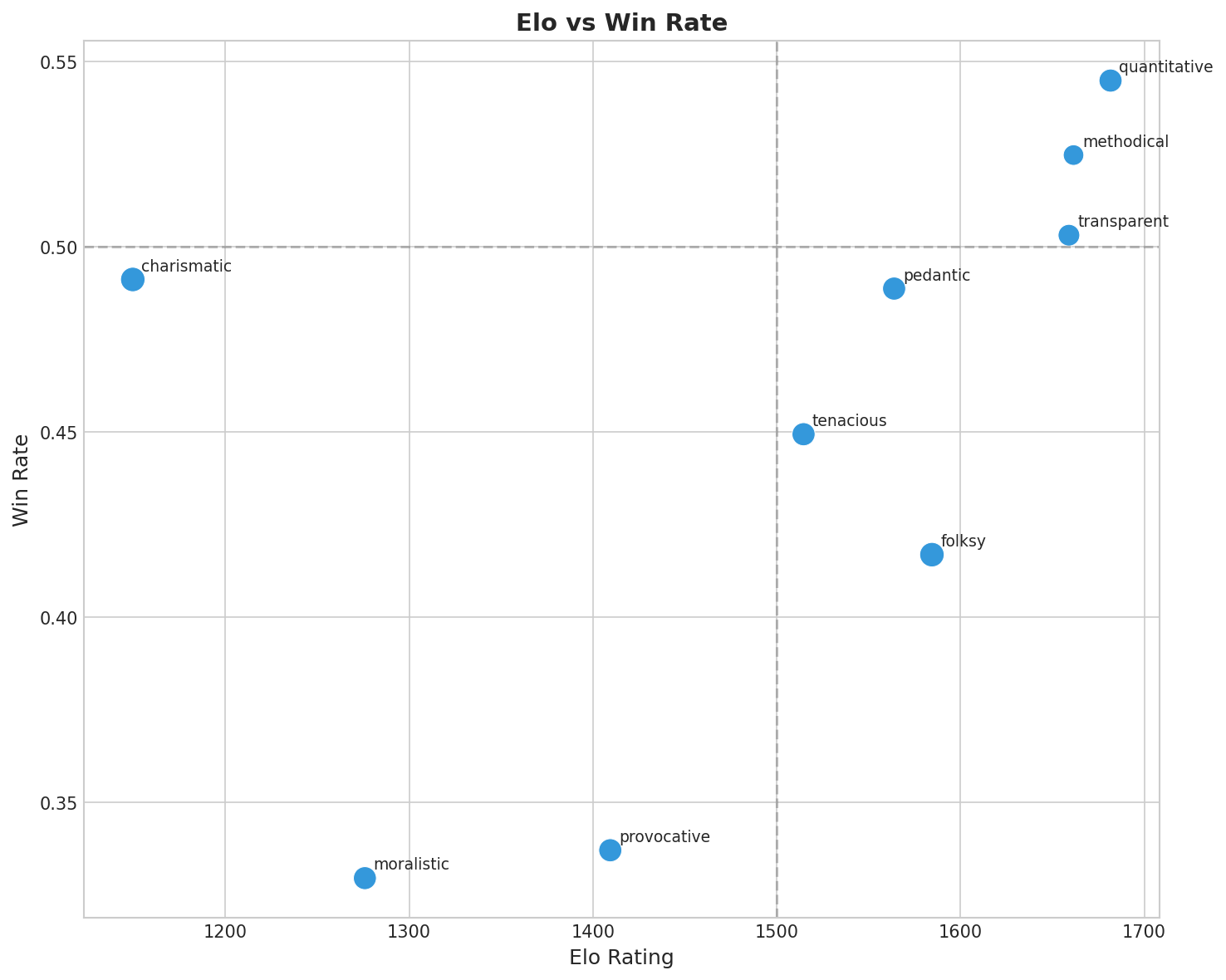}
\caption{Relationship between trait Elo rating and win rate under Single Agent, 2 Traits, 3 Rounds using Gemini-2.5-Pro. Each point represents a trait.}
\end{minipage}
\hspace{0.1cm}
\begin{minipage}{0.3\linewidth}
\centering
\includegraphics[width=\linewidth]{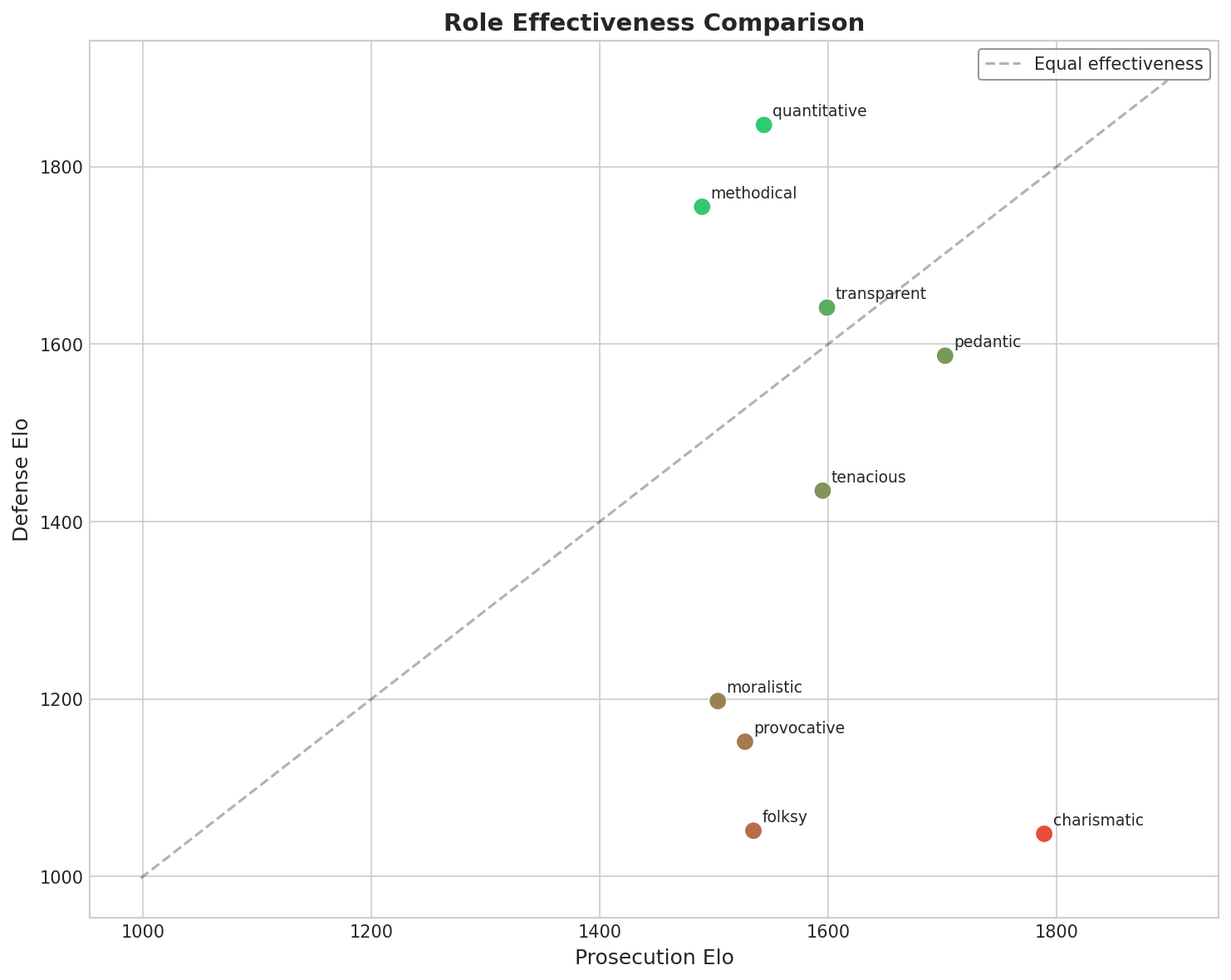}
\caption{Comparison of trait effectiveness across agent roles under Single Agent, 2 Traits, 3 Rounds using Gemini-2.5-Pro, showing prosecution versus defense Elo for each trait.}
\end{minipage}
\end{figure}

\newpage

\subsection{Team - 1 Trait - 2 Rounds - DeepSeek-R1}

\begin{figure}[H]
    \centering
    \includegraphics[width=0.63\linewidth]{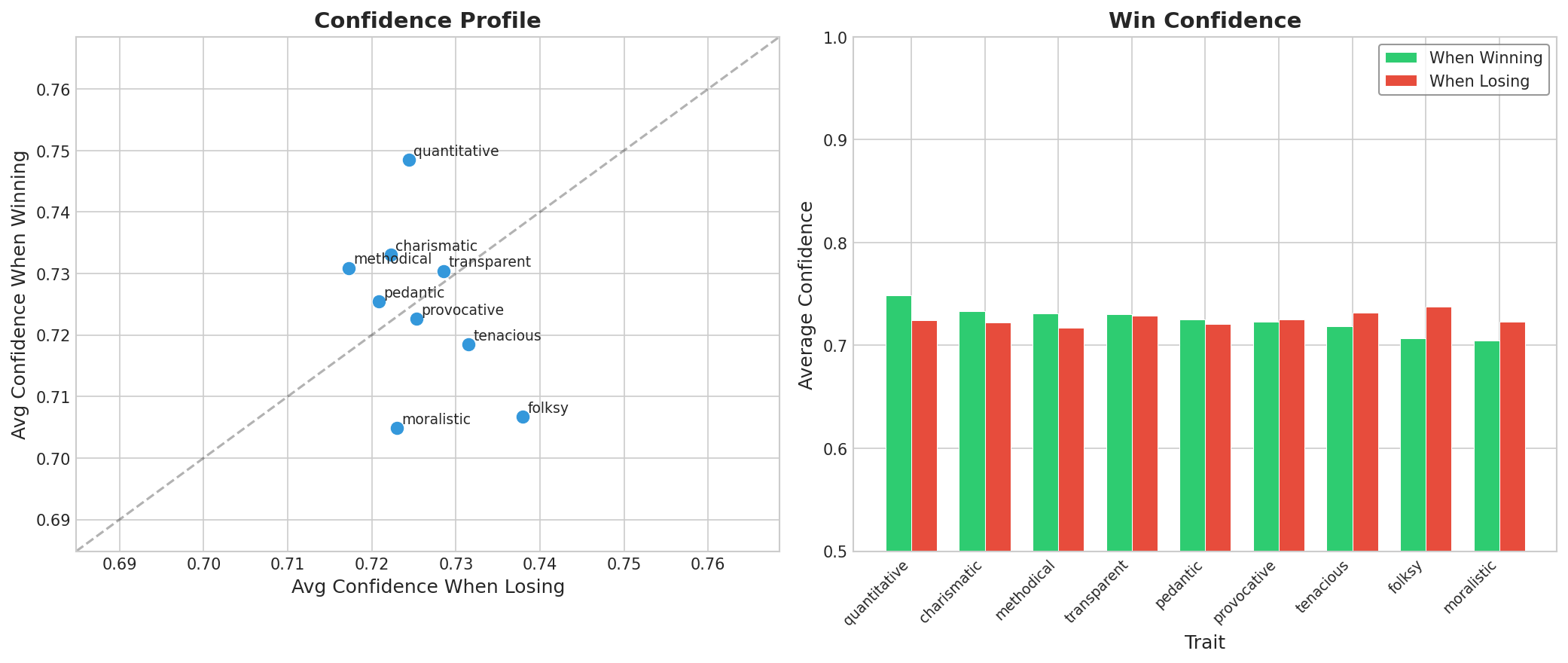}
    \caption{Trait confidence analysis for Team, 1 Trait, 2 Rounds using DeepSeek-R1. Left: Average judge confidence when a trait wins versus loses. Right: Mean confidence across traits for winning and losing cases.}
\end{figure}

\begin{figure}[H]
    \centering
    \includegraphics[width=0.51\linewidth]{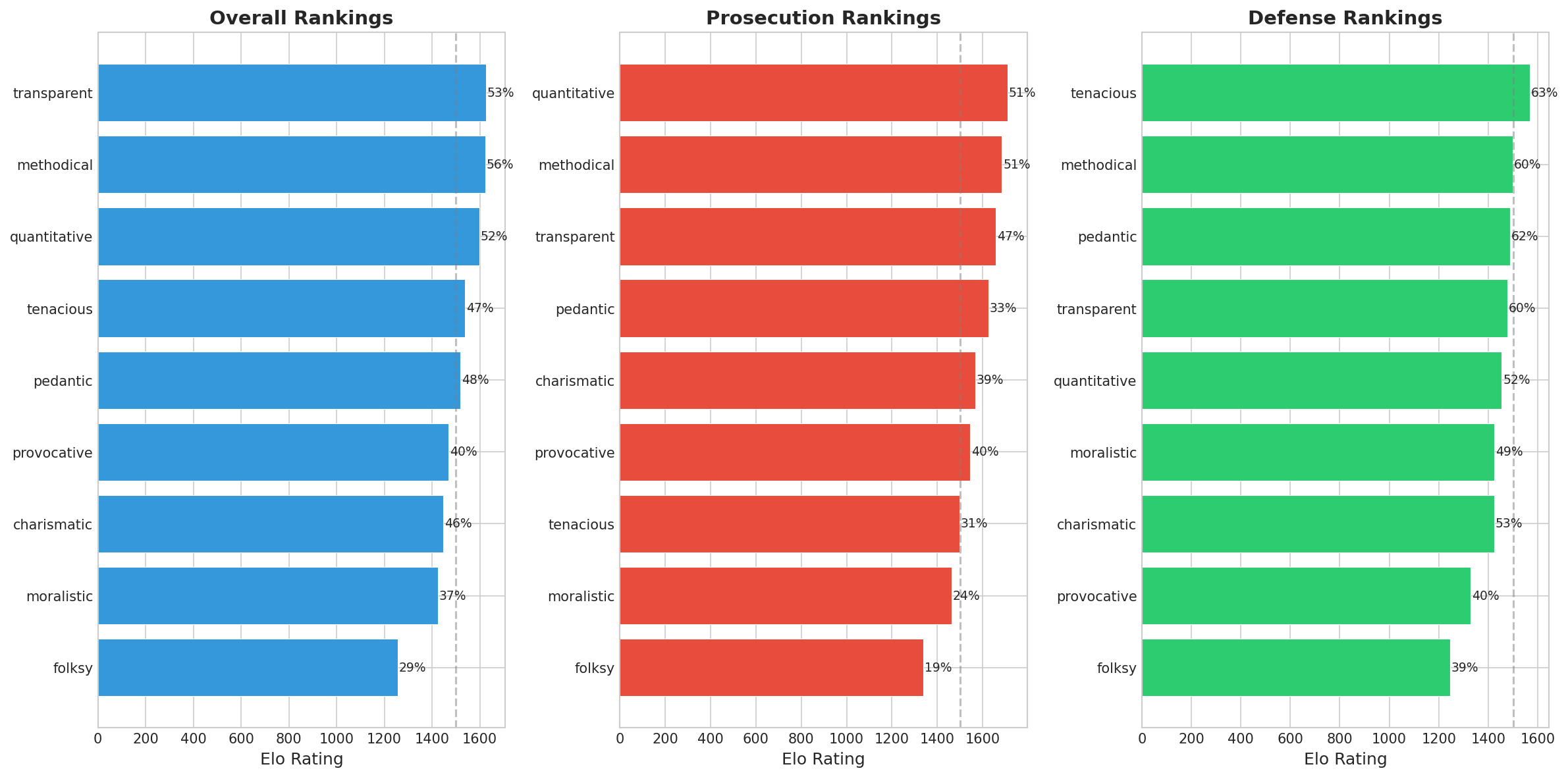}
    \caption{Elo rankings of traits under Team, 1 Trait, 2 Rounds using DeepSeek-R1, shown for overall performance, prosecution role, and defense role.}
\end{figure}

\begin{figure}[H]
    \centering
    \begin{minipage}{0.3\linewidth}
        \centering
        \includegraphics[width=\linewidth]{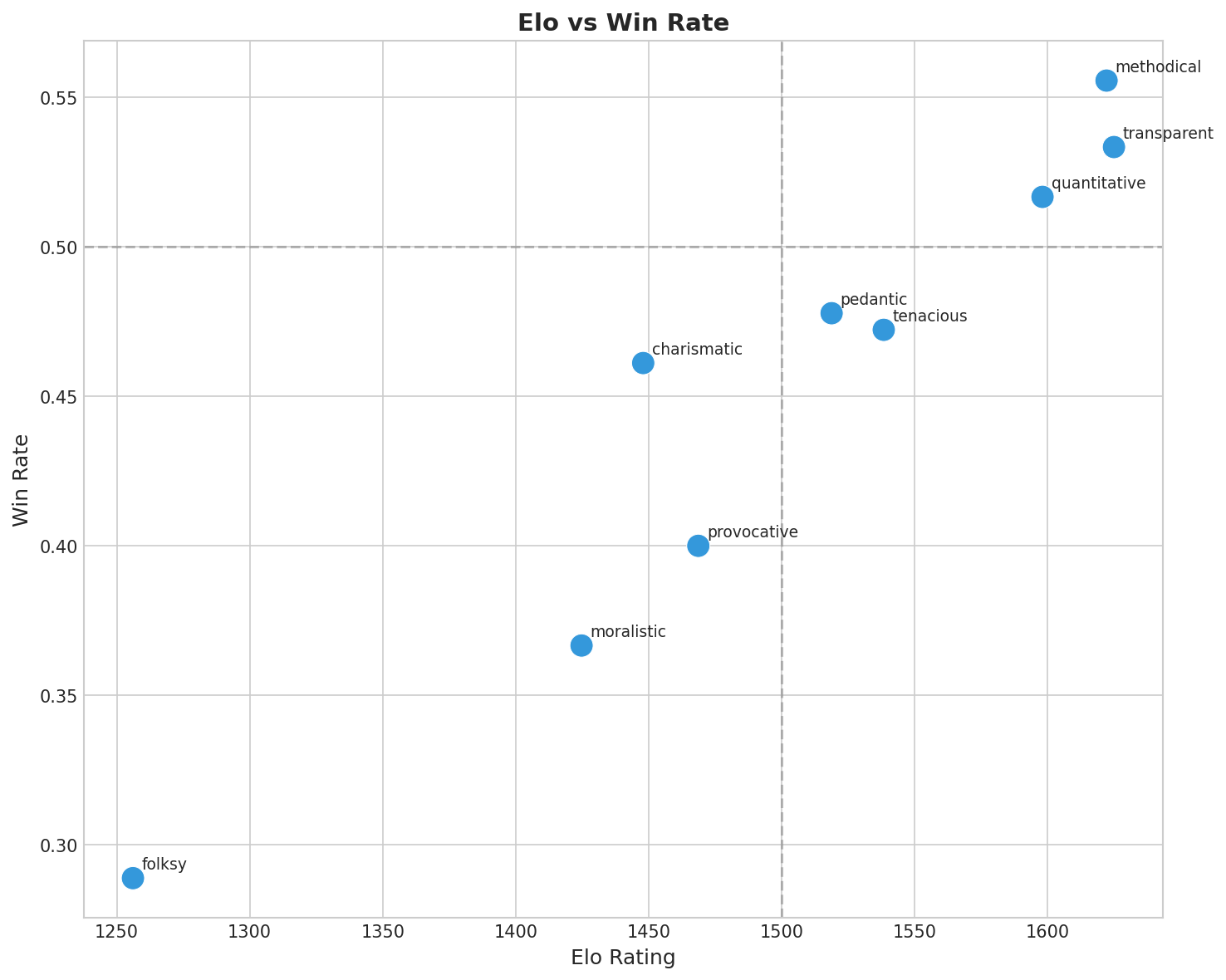}
        \caption{Relationship between trait Elo rating and win rate under Team, 1 Trait, 2 Rounds using DeepSeek-R1. Each point represents a trait.}
    \end{minipage}
    \hspace{0.1cm}
    \begin{minipage}{0.3\linewidth}
        \centering
        \includegraphics[width=\linewidth]{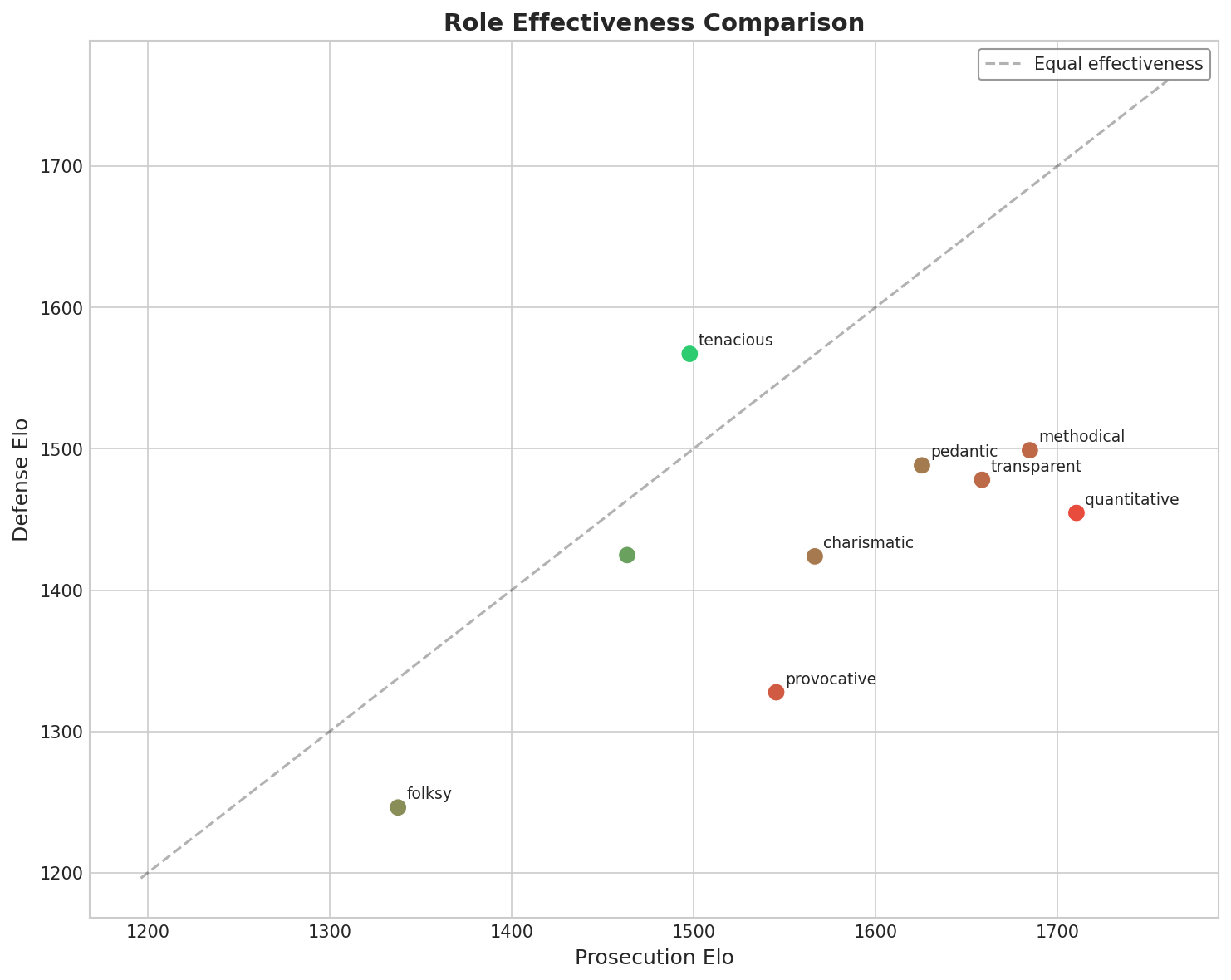}
        \caption{Comparison of trait effectiveness across agent roles under Team, 1 Trait, 2 Rounds using DeepSeek-R1, showing prosecution versus defense Elo for each trait.}
    \end{minipage}
\end{figure}

\newpage

\subsection{Team - 1 Trait - 3 Rounds - DeepSeek-R1}

\begin{figure}[H]
    \centering
    \includegraphics[width=0.63\linewidth]{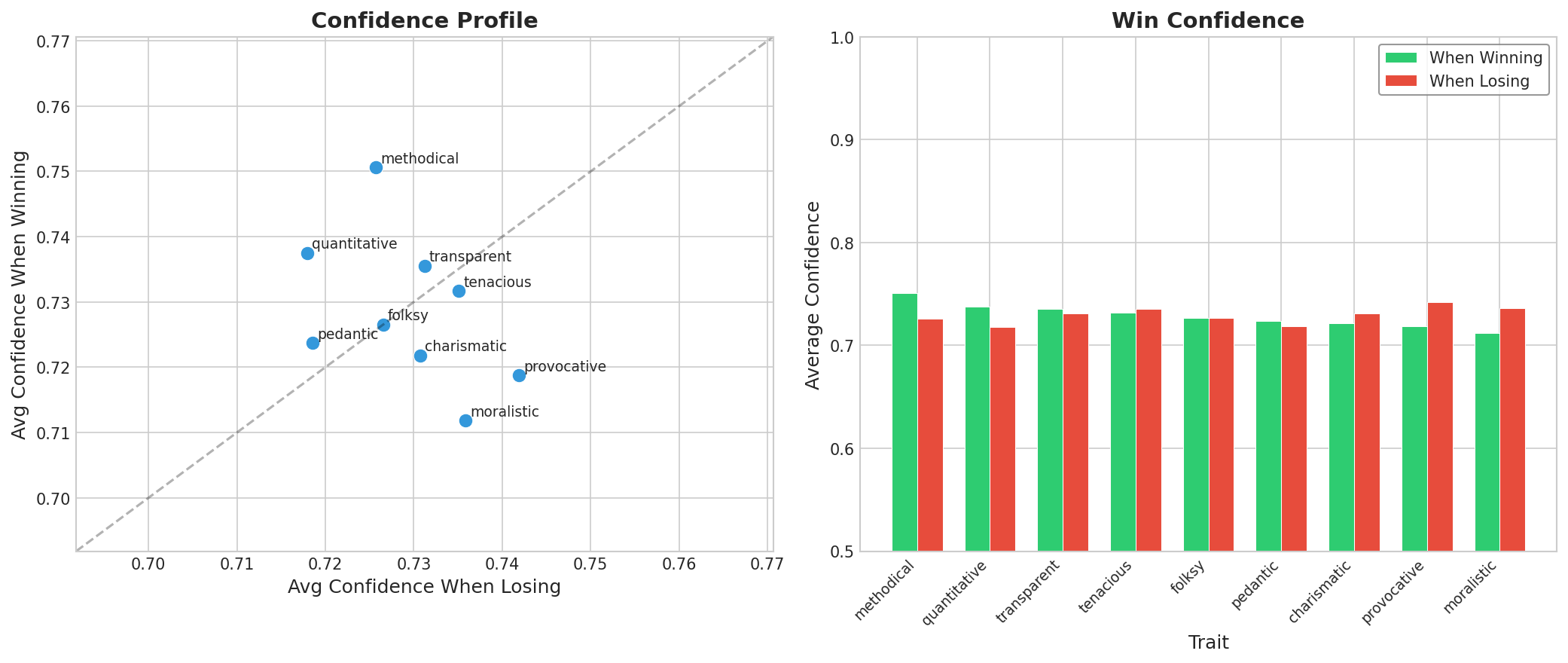}
    \caption{Trait confidence analysis for Team, 1 Trait, 3 Rounds using DeepSeek-R1. Left: Average judge confidence when a trait wins versus loses. Right: Mean confidence across traits for winning and losing cases.}
\end{figure}

\begin{figure}[H]
    \centering
    \includegraphics[width=0.51\linewidth]{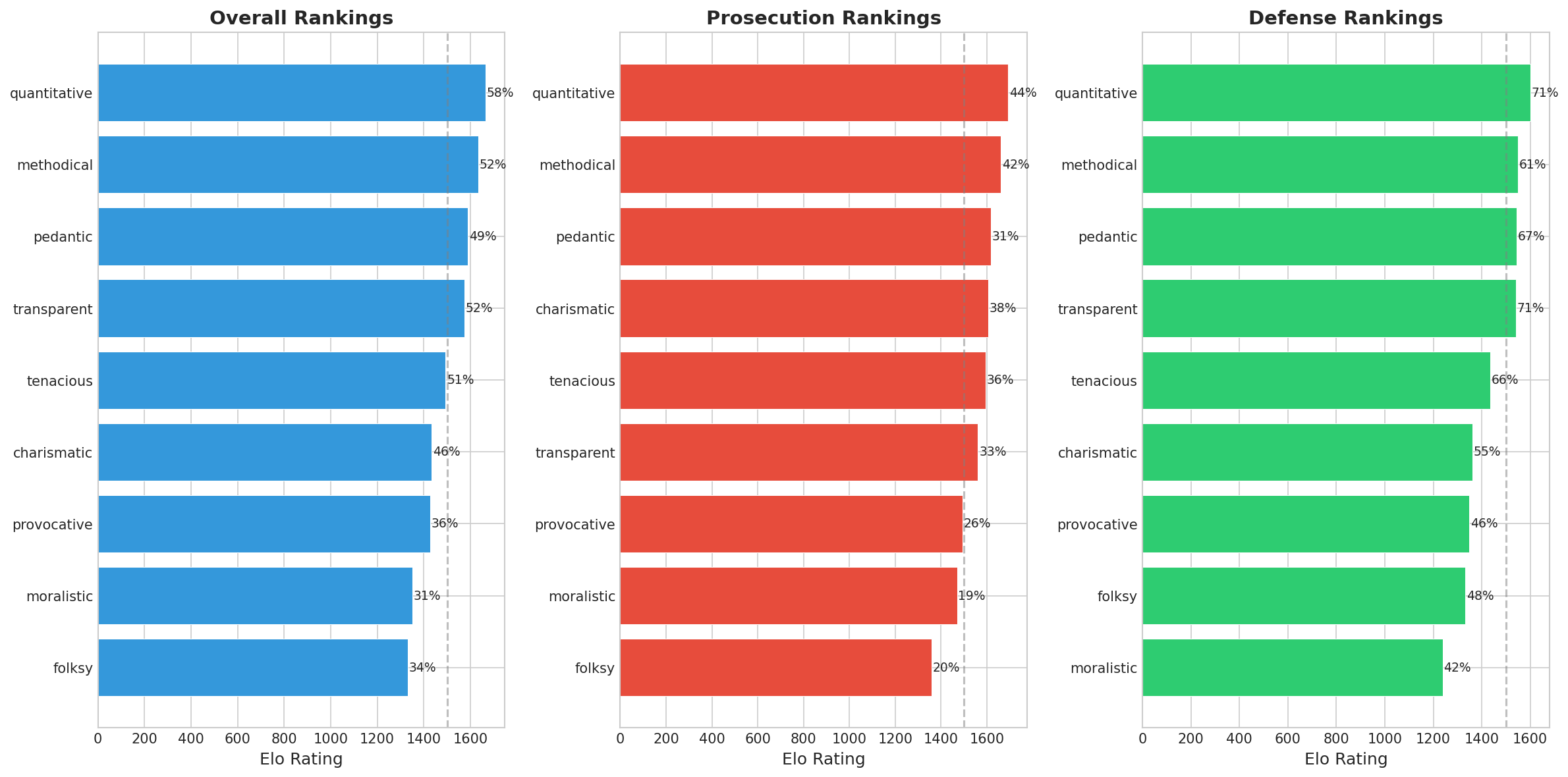}
    \caption{Elo rankings of traits under Team, 1 Trait, 3 Rounds using DeepSeek-R1, shown for overall performance, prosecution role, and defense role.}
\end{figure}

\begin{figure}[H]
    \centering
    \begin{minipage}{0.3\linewidth}
        \centering
        \includegraphics[width=\linewidth]{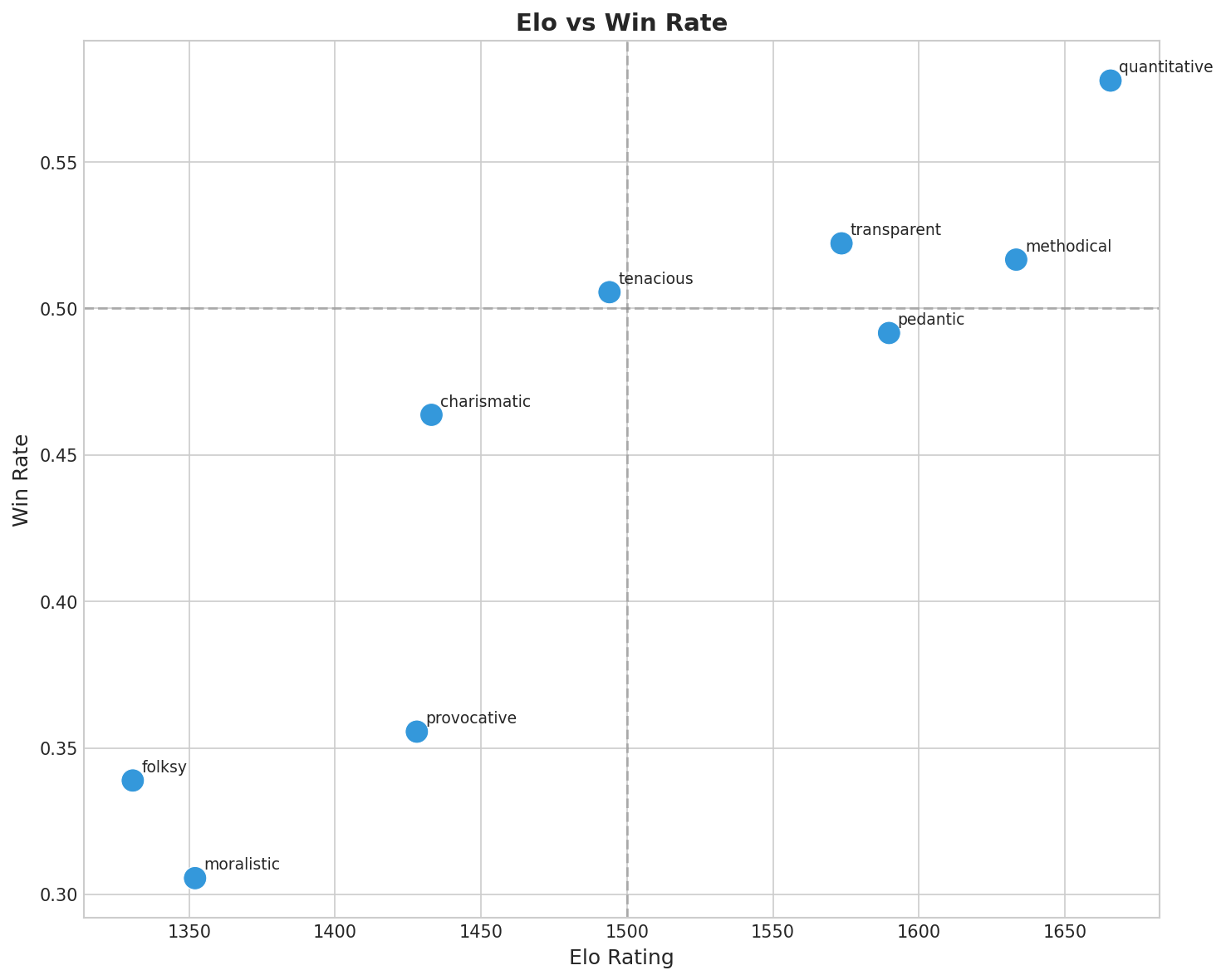}
        \caption{Relationship between trait Elo rating and win rate under Team, 1 Trait, 3 Rounds using DeepSeek-R1. Each point represents a trait.}
    \end{minipage}
    \hspace{0.1cm}
    \begin{minipage}{0.3\linewidth}
        \centering
        \includegraphics[width=\linewidth]{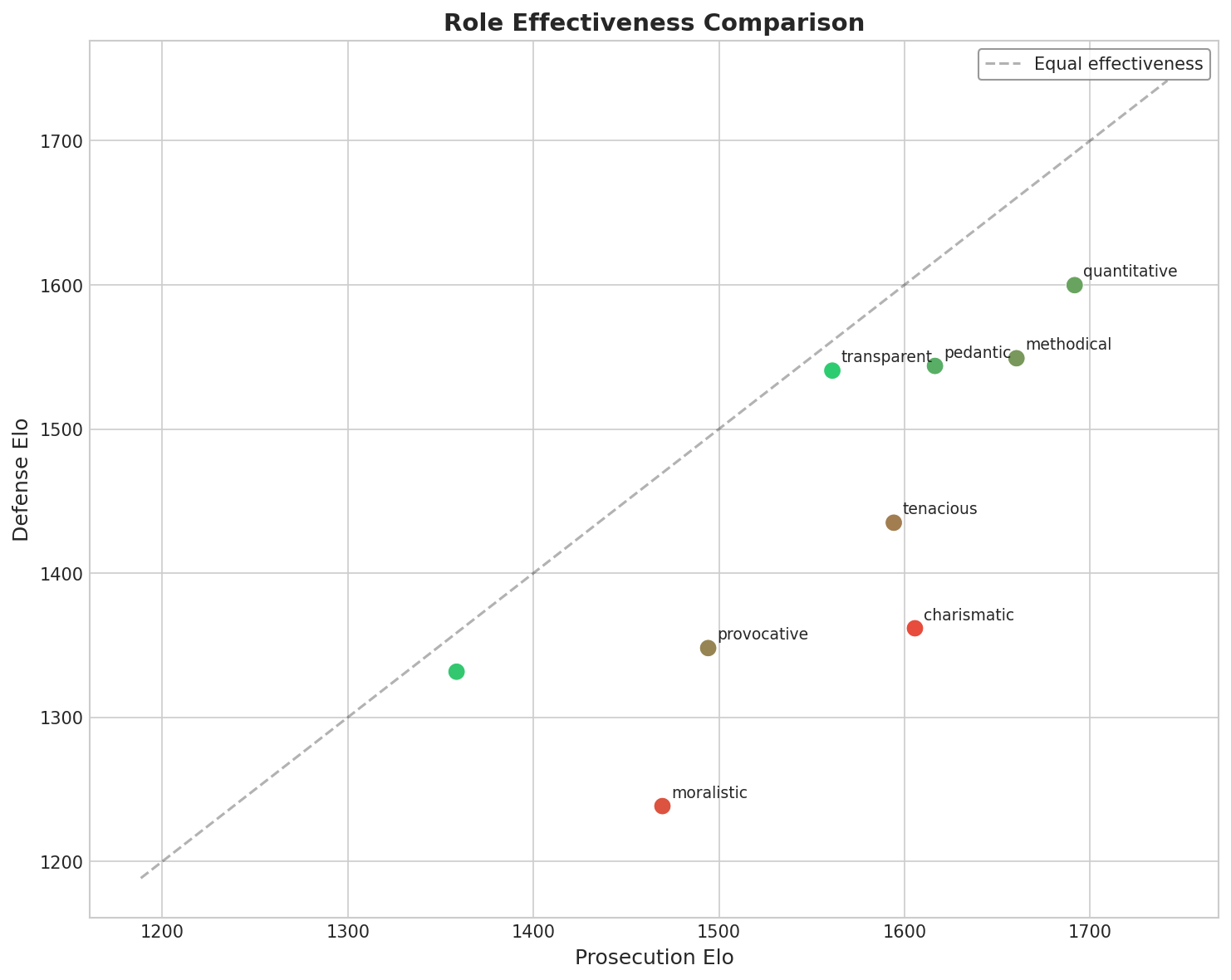}
        \caption{Comparison of trait effectiveness across agent roles under Team, 1 Trait, 3 Rounds using DeepSeek-R1, showing prosecution versus defense Elo for each trait.}
    \end{minipage}
\end{figure}

\newpage

\subsection{Team - 2 Traits - 3 Rounds - Gemini-2.5-Pro}

\begin{figure}[H]
    \centering
    \includegraphics[width=0.63\linewidth]{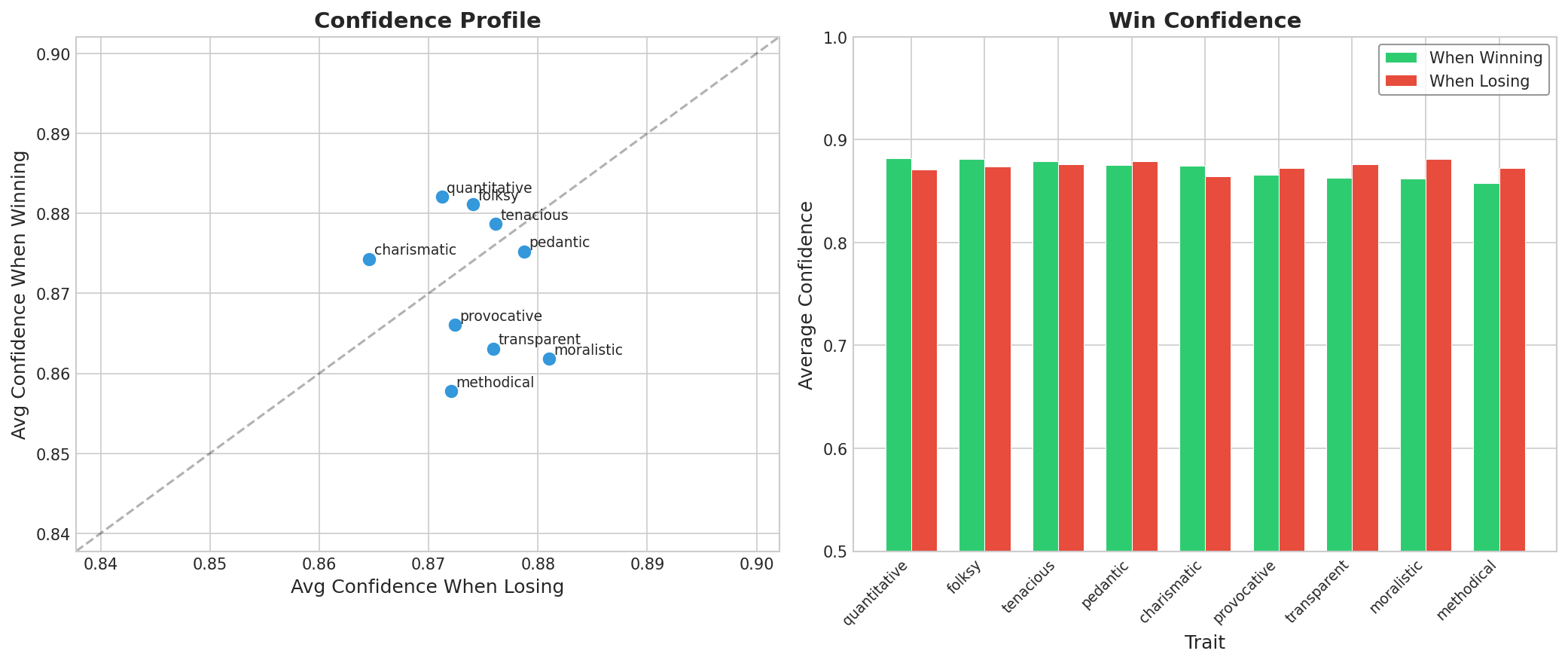}
    \caption{Trait confidence analysis for Team, 2 Traits, 3 Rounds using Gemini-2.5-Pro. Left: Average judge confidence when a trait wins versus loses. Right: Mean confidence across traits for winning and losing cases.}
\end{figure}

\begin{figure}[H]
    \centering
    \includegraphics[width=0.51\linewidth]{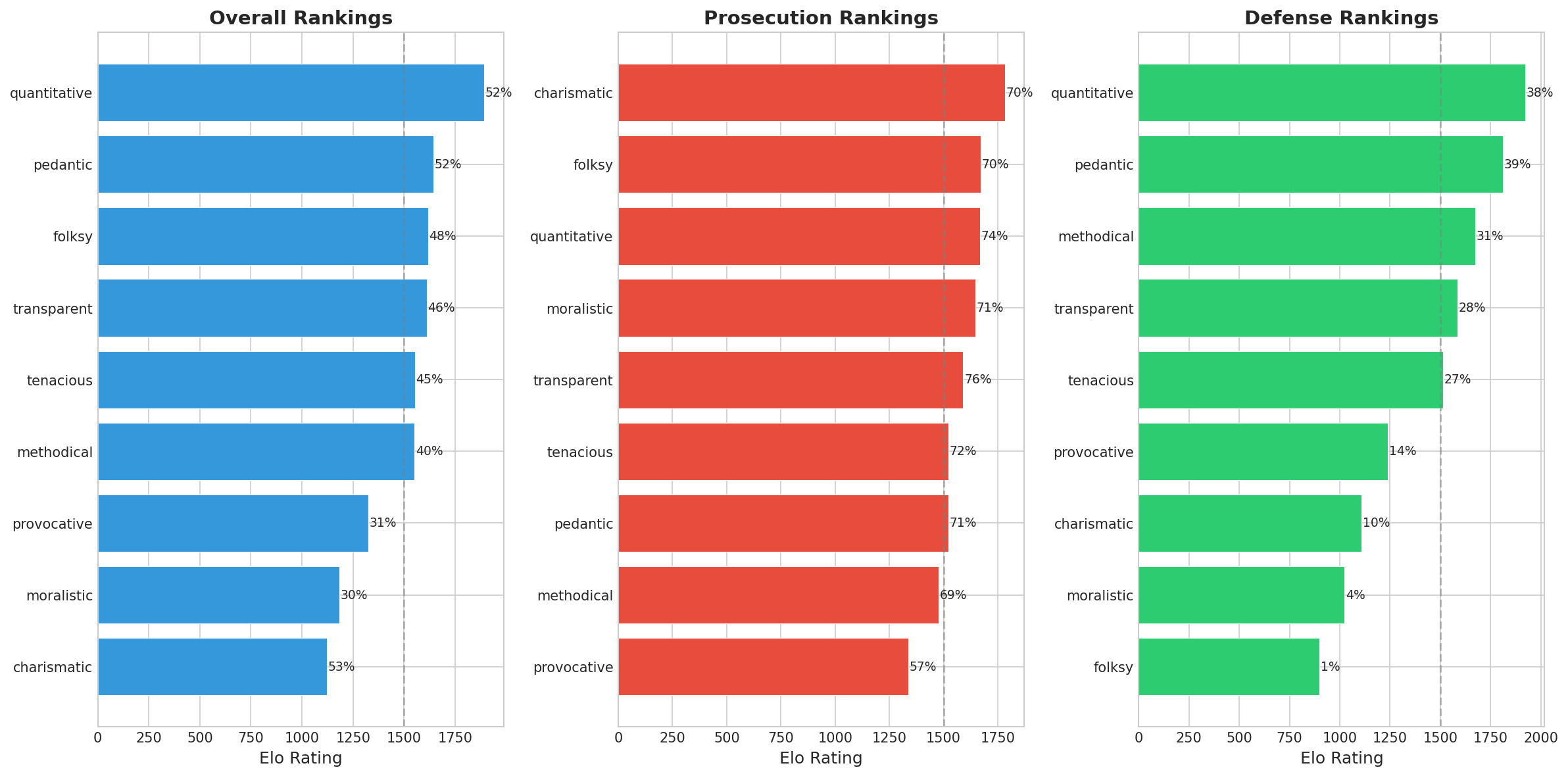}
    \caption{Elo rankings of traits under Team, 2 Traits, 3 Rounds using Gemini-2.5-Pro, shown for overall performance, prosecution role, and defense role.}
\end{figure}

\begin{figure}[H]
    \centering
    \begin{minipage}{0.3\linewidth}
        \centering
        \includegraphics[width=\linewidth]{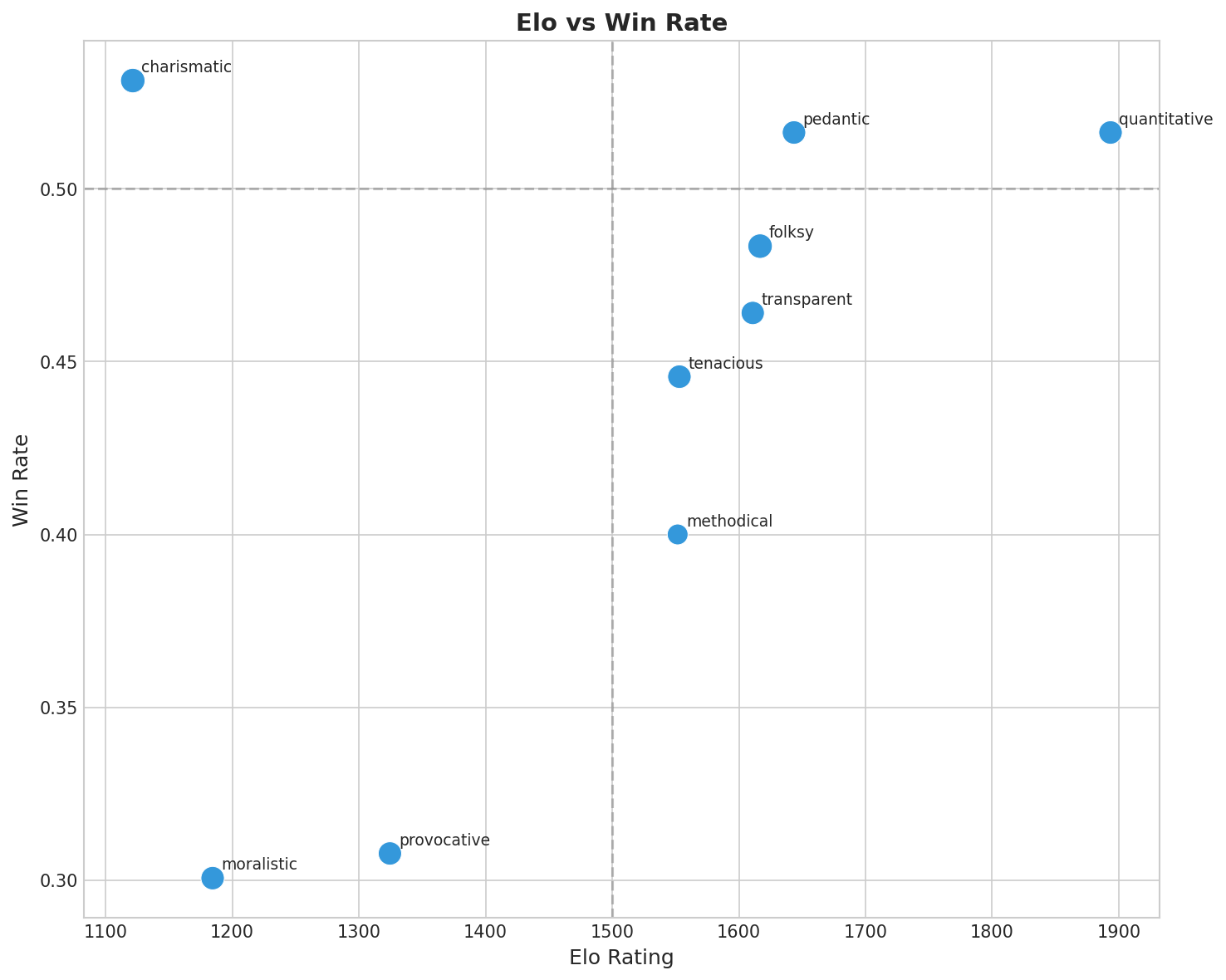}
        \caption{Relationship between trait Elo rating and win rate under Team, 2 Traits, 3 Rounds using Gemini-2.5-Pro. Each point represents a trait.}
    \end{minipage}
    \hspace{0.1cm}
    \begin{minipage}{0.3\linewidth}
        \centering
        \includegraphics[width=\linewidth]{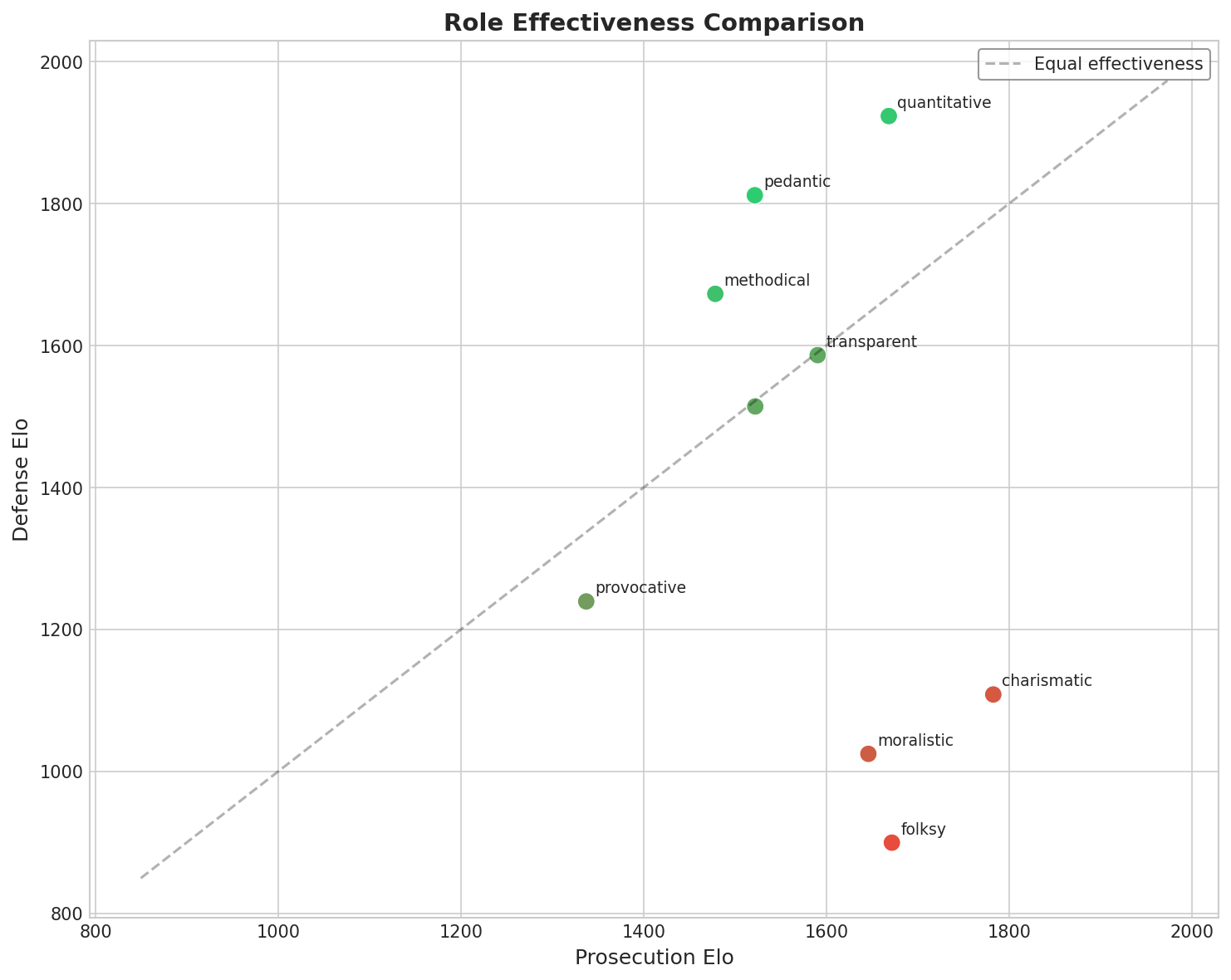}
        \caption{Comparison of trait effectiveness across agent roles under Team, 2 Traits, 3 Rounds using Gemini-2.5-Pro, showing prosecution versus defense Elo for each trait.}
    \end{minipage}
\end{figure}

\end{document}